\documentclass[journal]{IEEEtran}
\usepackage{cite}
\usepackage{graphicx}
\usepackage{amssymb,amsmath,bm}
\usepackage[colorlinks,linkcolor=black,anchorcolor=black,citecolor=black,urlcolor=black]{hyperref}
\usepackage{hyperref}
\usepackage{multirow}
\usepackage{amsfonts}
\usepackage{url}
\usepackage{multirow}
\usepackage{color}
\usepackage{tabularx}
\usepackage{color}
\ifCLASSINFOpdf
\else
\fi
\begin{document}
%\title{Articulatory-to-Acoustic Conversion Using Bidirectional Long Short-Term Memory Based Recurrent Neural Networks with Augmented Input Representation}
\title{Low-Latency Neural Speech Phase Prediction based on Parallel Estimation Architecture and Anti-Wrapping Losses for Speech Generation Tasks}

\author{Yang~Ai,~\IEEEmembership{Member,~IEEE},~Zhen-Hua~Ling,~\IEEEmembership{Senior Member,~IEEE}% <-this % stops a space
\thanks{This work is the extended version of our conference paper \cite{ai2023neural} published at 2023 IEEE International Conference on Acoustics, Speech and Signal Processing (ICASSP 2023).}
\thanks{This work was funded by the National Nature Science Foundation of China under Grant 62301521 and U23B2053, the Anhui Provincial Natural Science Foundation under Grant 2308085QF200, and the Fundamental Research Funds for the Central Universities under Grant WK2100000033.}
%\thanks{Y. Ai is with the National University of Defense Technology, Hefei, 230037, China (e-mail: ay8067@ustc.edu). This work was done when he was a graduate student at the National Engineering Laboratory for Speech and Language Information Processing, University of Science and Technology of China.}
\thanks{Y. Ai and Z.-H. Ling are with the National Engineering Research Center of Speech and Language Information Processing, University of Science and Technology of China, Hefei, 230027, China (e-mail: yangai@ustc.edu.cn, zhling@ustc.edu.cn).}
\thanks{Corresponding author: Zhen-Hua Ling.}
%\thanks{H.-Y. Li, X. Wang and J. Yamagishi are with with the National Institute of Informatics, Tokyo 101-8340, Japan (e-mail: \{haoyuli,wangxin,jyamagis\}@nii.ac.jp).}
%\thanks{Y. Gu is with Baidu Speech Department, Baidu Technology Park, Beijing, 100193, China (e-mail: guyu04@baidu.com ). This work was done when he was
%a graduate student at the National Engineering Laboratory of Speech and Language Information Processing,
%University of Science and Technology of China. }
%}
%\thanks{Manuscript received April 19, 2005; revised January 11, 2007.}}
}

\markboth{}%
{Shell \MakeLowercase{\textit{et al.}}: Bare Demo of IEEEtran.cls for Journals}



\maketitle

\begin{abstract}
%\boldmath
This paper presents a novel neural speech phase prediction model which predicts wrapped phase spectra directly from amplitude spectra.
The proposed model is a cascade of a residual convolutional network and a parallel estimation architecture.
The parallel estimation architecture is a core module for direct wrapped phase prediction.
This architecture consists of two parallel linear convolutional layers and a phase calculation formula, imitating the process of calculating the phase spectra from the real and imaginary parts of complex spectra and strictly restricting the predicted phase values to the principal value interval.
To avoid the error expansion issue caused by phase wrapping, we design anti-wrapping training losses defined between the predicted wrapped phase spectra and natural ones by activating the instantaneous phase error, group delay error and instantaneous angular frequency error using an anti-wrapping function.
We mathematically demonstrate that the anti-wrapping function should possess three properties, namely parity, periodicity and monotonicity.
We also achieve low-latency streamable phase prediction by combining causal convolutions and knowledge distillation training strategies.
For both analysis-synthesis and specific speech generation tasks, experimental results show that our proposed neural speech phase prediction model outperforms the iterative phase estimation algorithms and neural network-based phase prediction methods in terms of phase prediction precision, efficiency and robustness.
Compared with HiFi-GAN-based waveform reconstruction method, our proposed model also shows outstanding efficiency advantages while ensuring the quality of synthesized speech.
%Experimental results show that our proposed neural speech phase prediction model outperforms the iterative Griffin-Lim algorithm, iterative relaxed averaged alternating reflection algorithm, von Mises distribution deep neural network-based phase prediction method and HiFi-GAN-based waveform reconstruction method for both analysis-synthesis and specific speech generation tasks in terms of phase prediction precision, efficiency, latency and robustness.
%Our proposed model is also more robust than iterative algorithms when using the degraded amplitude spectra as input for specific speech generation tasks.
To the best of our knowledge, we are the first to directly predict speech phase spectra from amplitude spectra only via neural networks.

\end{abstract}


% \fi
\IEEEpeerreviewmaketitle

\section{Introduction}

\IEEEPARstart{S}peech phase prediction, also known as speech phase reconstruction, recovers speech phase spectra from amplitude spectra and plays an important role in speech generation tasks.
Currently, several speech generation tasks, such as speech enhancement (SE) \cite{lu2013speech,xu2014regression,kim2020t}, bandwidth extension (BWE) \cite{wang2015speech,gu2016speech,li2015dnn} and speech synthesis (SS) \cite{zen2009statistical,wang2017tacotron,shen2018natural,takaki2017direct,marafioti2019adversarial,saito2018text}, mainly focus on the prediction of amplitude spectra or amplitude-derived features (e.g., mel spectrograms and mel cepstra).
Therefore, speech phase prediction is crucial for waveform reconstruction in these tasks.
However, limited by the issue of phase wrapping and the difficulty of phase modeling, the precise prediction of the speech phase remains a challenge until now.
In addition to phase prediction precision, the efficiency, latency and robustness are also important metrics for evaluating phase prediction methods.
The efficiency represents the generation speed and is a criterion for determining real-time performance.
The latency refers to the duration of future input that are necessary for predicting current output.
High efficiency and low latency are strict requirements for many practical application scenarios such as telecommunication.
The robustness reflects the general applicability of phase prediction methods when faced with inputs of varying amplitude spectra.

In the early days, researchers mainly focused on iterative phase estimation algorithms, of which the Griffin-Lim algorithm (GLA) \cite{griffin1984signal} is one of the most well-known algorithms.
The GLA is based on alternating projection and iteratively estimates the phase spectra from amplitude spectra via the short-time Fourier transform (STFT) and inverse STFT (ISTFT).
Due to its ease of implementation, the GLA has been widely used in speech generation tasks \cite{takaki2017direct,marafioti2019adversarial,saito2018text}.
However, the GLA always causes unnatural artifacts in the reconstructed speech, meaning that there is still a large gap between the estimated phase and natural phase.
Hence, several improved algorithms, such as the fast Griffin-Lim algorithm (FGLA) \cite{perraudin2013fast} and alternating direction method of multipliers (ADMM) \cite{masuyama2018griffin}, have also been also proposed to boost the performance of the GLA.
Recently, Kobayashi \MakeLowercase{\textit{et al.}} \cite{kobayashi2022acoustic} applied three alternating reflection-based iterative algorithms developed in the optics community, i.e., the averaged alternating reflections (AAR) \cite{bauschke2002phase,bauschke2004finding}, relaxed AAR (RAAR) \cite{luke2004relaxed}, and hybrid input-output (HIO) \cite{fienup1978reconstruction,fienup1982phase,bauschke2003hybrid} algorithms, to acoustic applications and clearly outperformed GLA families.
However, these iterative algorithms always limit the reconstructed speech quality due to the influence of the initial phase and exhibit poor robustness when estimating phase spectra from degraded amplitude spectra.
Besides, these iterative algorithms often require the amplitude spectra of an entire utterance as input, resulting in high latency.
%Besides, Some application scenarios have strict requirement on the latency, which refers to the duration of future input that are necessary for predicting current output.

With the development of deep learning, several neural network-based phase prediction methods have been gradually proposed.
We roughly divide them into three categories of methods.
The first is the GLA simulation method \cite{oyamada2018generative,masuyama2019deep,masuyama2020deep}.
For example, Masuyama \MakeLowercase{\textit{et al.}} \cite{masuyama2019deep,masuyama2020deep} proposed deep Griffin-Lim iteration (DeGLI), which employs trainable neural networks to simulate the GLA process and achieve iterative phase reconstruction.
However, the prediction target of such methods is the complex spectra rather than the phase spectra.
The second is the two-stage method \cite{masuyama2020phase,thieling2021recurrent,thien2021two}.
For example, Masuyama \MakeLowercase{\textit{et al.}} \cite{masuyama2020phase} first predicted phase derivatives (i.e., the group delay and instantaneous frequency) by two parallel deep neural networks (DNNs), and then the phase was recursively estimated by a recurrent phase unwrapping (RPU) algorithm \cite{ghiglia1998two} from the predicted phase derivatives.
Prior-distribution-aware method is the last category \cite{takamichi2018phase,takamichi2020phase}.
Takamichi \MakeLowercase{\textit{et al.}} \cite{takamichi2018phase,takamichi2020phase} assume that the phase follows a specific prior distribution (i.e., the von Mises distribution and sine-skewed generalized cardioid distribution) and employ a DNN to predict the distribution parameters of the phase.
However, the phase predicted by the DNN still needs to be refined using the GLA.
Obviously, all the abovementioned phase prediction methods need to combine neural networks with some convolutional iterative algorithms, which inevitably leads to cumbersome operations, increased complexity, low efficiency and high latency.

%Vocoders \cite{dudley1939vocoder}, which reconstruct speech waveforms from acoustic features (e.g., mel spectrograms), can also be used for phase prediction when using amplitude spectra as input.
%Vocoder-based phase prediction methods are an implicit and indirect approach that includes phase prediction within waveform synthesis.
%Recently, HiFi-GAN \cite{kong2020hifi} vocoder has demonstrated exceptional performance on reconstructed speeches and has been widely applied in speech synthesis.
%However, direct waveform prediction imposes constraints on the generation efficiency of HiFi-GAN, while the extensive use of non-causal convolutions contributes to increased latency.
%The training strategy employed by the generative adversarial network (GAN) \cite{goodfellow2014generative} also reduces the training speed.
%To our knowledge, predicting speech wrapped phase spectra directly from amplitude spectra using only neural networks has not yet been thoroughly investigated.

In addition to the phase prediction methods mentioned above, several speech waveform reconstruction methods, such as vocoders \cite{dudley1939vocoder}, also include implicit and indirect phase prediction within waveform synthesis.
Recently, HiFi-GAN \cite{kong2020hifi} vocoder has demonstrated exceptional performance on reconstructed waveform and has been widely applied in speech synthesis.
The HiFi-GAN cascades multiple upsampling layers and residual convolution networks to gradually upsample the input mel spectrograms to the sampling rate of the final waveform while performing non-causal residual convolutional operations.
Hence, the phase prediction is implicitly incorporated within waveform prediction.
The HiFi-GAN utilizes adversarial losses \cite{goodfellow2014generative} defined on the waveform, ensuring the generation of high-fidelity waveforms.
However, HiFi-GAN still has limitations in terms of generation efficiency, training efficiency and latency, due to the direct waveform prediction, adversarial training and non-causal convolutions, respectively.
%However, direct waveform prediction imposes constraints on the generation efficiency of HiFi-GAN, while the extensive use of non-causal convolutions contributes to increased latency.
%The training strategy employed by the generative adversarial network (GAN) \cite{goodfellow2014generative} also reduces the training speed.
To our knowledge, predicting speech wrapped phase spectra directly from amplitude spectra using only neural networks has not yet been thoroughly investigated.

Due to the phase wrapping property, how to design 1) suitable architectures or activation functions to restrict the range of predicted phases for direct wrapped phase prediction and 2) loss functions suitable for phase characteristics, are the two major challenges for direct phase prediction based on neural networks.
To overcome these challenges, we propose a neural speech phase prediction model based on a parallel estimation architecture and anti-wrapping losses.
The proposed model passes the input log amplitude spectra through a residual convolutional network and a parallel estimation architecture to predict the wrapped phase spectra directly.
To restrict the output phase values to the principal value interval and predict the wrapped phases directly, the parallel estimation architecture imitates the process of calculating the phase spectra from the real and imaginary parts of complex spectra, and it is formed by two parallel convolutional layers and a phase calculation formula.
Due to the periodic nature and wrapping property of the phase, some conventional loss functions, such as L1 loss and mean square error (MSE), are disabled for phase prediction and cause error expansion issue.
To avoid the error expansion issue caused by phase wrapping, we propose the instantaneous phase loss, group delay loss and instantaneous angular frequency loss activated by an anti-wrapping function at the training stage.
These losses are defined between the phase spectra predicted by the model and the natural ones.
The anti-wrapping function calculates the true error between the predicted value and the natural value, and we demonstrate that the function requires three properties, i.e., parity, periodicity, and monotonicity.
We have also employed knowledge distillation to train an all-causal convolution-based neural phase prediction model.
%This approach has enabled us to achieve precise phase prediction with zero latency, thereby facilitating its effective utilization in low-latency scenarios.
This approach has enabled us to achieve precise streamable phase prediction, thereby facilitating its effective utilization in low-latency scenarios.
Experimental results show that our proposed model outperforms the GLA \cite{griffin1984signal}, RAAR \cite{kobayashi2022acoustic} and von Mises distribution DNN-based phase prediction method \cite{takamichi2018phase,takamichi2020phase}, in terms of both phase prediction precision and efficiency.
Compared with the HiFi-GAN vocoder-based waveform reconstruction method \cite{kong2020hifi}, our proposed model demonstrates a significant efficiency advantage while maintaining the same quality of synthesized speech.
%and HiFi-GAN vocoder-based waveform reconstruction method \cite{kong2020hifi} by comprehensive  the phase prediction precision, efficiency, latency and robustness.
Our proposed model exhibits near-natural reconstructed speech quality according to the mean opinion score (MOS) results and reaches 19.6x real-time generation on a CPU with low latency.
When the proposed method is initially applied to specific speech generation tasks (i.e., using degraded amplitude spectra as input), it shows better stability and robustness than iterative algorithms.
Ablation studies also certify that the parallel estimation architecture and anti-wrapping losses are extremely important for successful phase prediction.

The main contribution of this work is the realization of direct prediction of the wrapped phase spectra only by neural networks with high prediction precision, high generation efficiency, low latency and high robustness.
%Compared with conventional iterative phase estimation algorithms, our proposed neural phase prediction model exhibits higher reconstructed speech quality, higher generation efficiency and higher robustness.
%Compared with existing neural network-based phase prediction methods, our proposed neural phase prediction model is easy to implement, simple to operate, and requires no extra algorithm support.
Our proposed model is easy to implement, simple to operate and adaptable to integrate into specific speech generation tasks, such as SE, BWE and SS, due to its trainable property.

This paper is organized as follows.
In Section \ref{sec: Related works}, we briefly review the representative iterative speech phase estimation algorithms and neural network-based speech phase prediction methods.
In Section \ref{sec: Proposed Methods}, we provide details of the model structure, training criteria and improvements made for low-latency streamable inference of the proposed neural speech phase prediction model.
In Section \ref{sec: Experiments}, we present our experimental results.
Finally, we give conclusions in Section \ref{sec: Conclusion}.

\section{Related Works}
\label{sec: Related works}

In this section, we briefly introduce two iterative speech phase estimation algorithms (i.e., the GLA \cite{griffin1984signal} and RAAR \cite{kobayashi2022acoustic}) and a neural network-based speech phase prediction method (i.e., the von Mises distribution DNN-based method \cite{takamichi2018phase,takamichi2020phase}).
They are compared with our proposed neural phase prediction model in Section \ref{sec: Experiments}.

\subsection{Griffin-Lim Algorithm (GLA)}
\label{subsec: Griffin-Lim algorithm (GLA)}

The GLA \cite{griffin1984signal} is an alternating projection algorithm and iteratively estimates the phase spectra from amplitude spectra via the STFT and ISTFT.
Assume that the amplitude spectrum is $\bm{A}\in \mathbb{R}^{F\times N}$, where $F$ and $N$ are the total number of frames and frequency bins, respectively.
Then initialize the phase spectrum $\hat{\bm{P}}^{[0]}\in \mathbb{R}^{F\times N}$ to zero matrix, i.e., the initial complex spectrum $\hat{\bm{S}}^{[0]}=\bm{A}\odot e^{j\hat{\bm{P}}^{[0]}}=\bm{A}$, where $\odot$ represents the element-wise multiplication.
Finally iterate the following formula from $i=1$ to $I$:
\begin{equation}
\label{equ: GL}
\hat{\bm{S}}^{[i]}=P_C(P_A(\hat{\bm{S}}^{[i-1]})),
\end{equation}
where $I$ is the total number of iterations.
$P_C$ and $P_A$ are two core projection operators defined as follows:
\begin{equation}
\label{equ: GL_PC}
P_C(\bm{X})=STFT(ISTFT(\bm{X})),
\end{equation}
\begin{equation}
\label{equ: GL_PA}
P_A(\bm{X})=\bm{A}\odot\bm{X}\oslash|\bm{X}|,
\end{equation}
where $\bm{X}\in \mathbb{C}^{F\times N}$.
$\oslash$ and $|\cdot|$ represent the element-wise division and amplitude calculation, respectively.
The final estimated phase spectrum $\hat{\bm{P}}^{[I]}\in \mathbb{R}^{F\times N}$ is contained in the complex spectrum $\hat{\bm{S}}^{[I]}\in \mathbb{C}^{F\times N}$.
The final speech waveform is reconstructed from $\hat{\bm{S}}^{[I]}$ by ISTFT.
The GLA can be easily implemented and is popular in speech generation tasks.
Since the GLA always gives a local optimal solution, the reconstructed speech quality is limited by the influence of the initial phase and there are obvious artifacts in the reconstructed speech.
Besides, the GLA also tends to limit the phase estimation efficiency due to its iterative estimation mode and extend the latency due to its whole-utterance estimation mode.

\subsection{Relaxed Averaged Alternating Reflection (RAAR)}
\label{subsec: Relaxed averaged alternating reflection (RAAR)}

The RAAR was originally developed in the optics community, and was recently successfully applied in the field of speech phase estimation by Kobayashi \MakeLowercase{\textit{et al.}} \cite{kobayashi2022acoustic}.
The RAAR is an alternating reflection algorithm and also iteratively estimates the phase spectra from the amplitude spectra.
The core reflection operators $R_C$ and $R_A$ for the RAAR are designed based on the projection operators $P_C$ and $P_A$ as follows:
\begin{equation}
\label{equ: RAAR_RC}
R_C(\bm{X})=2P_C(\bm{X})-\bm{X},
\end{equation}
\begin{equation}
\label{equ: RAAR_RA}
R_A(\bm{X})=2P_A(\bm{X})-\bm{X}.
\end{equation}
The RAAR adopts the same initialization manner as the GLA and then iteratively executes the following formula from $i=1$ to $I$:
\begin{equation}
\label{equ: RAAR}
\hat{\bm{S}}^{[i]}=\dfrac{\beta}{2}\hat{\bm{S}}^{[i-1]}+R_C(R_A(\hat{\bm{S}}^{[i-1]}))+(1-\beta)P_A(\hat{\bm{S}}^{[i-1]}),
\end{equation}
where $0<\beta<1$ is a relaxation parameter.

In the original paper \cite{kobayashi2022acoustic}, Kobayashi \MakeLowercase{\textit{et al.}} have proven that the RAAR with $\beta=0.9$ is an excellent speech phase estimation algorithm which outperforms the GLA families and other alternating reflection algorithms (i.e., the AAR and HIO).
However, the iterative formula of the RAAR is more complicated than that of the GLA, which inevitably inhibits the generation efficiency and latency.

\subsection{Von Mises Distribution DNN-based Method}
\label{subsec: Von-Mises-distribution DNN-based method}

The von Mises distribution DNN-based method \cite{takamichi2018phase,takamichi2020phase} realizes phase prediction by combining neural networks and GLA.
It assumes that the phase follows a von Mises distribution and then uses a DNN to predict the mean parameter of the phase distribution from the input log amplitude spectra at current and $\pm$2 frames.
The mean parameter is regarded as the predicted phase.
The DNN is composed of three 1024-unit feed-forward hidden layers activated by a gated linear unit (GLU) \cite{dauphin2017language} and a linear output layer.
A multi-task learning strategy with phase loss and group delay loss is adopted to train the DNN.
The phase loss and group delay loss are formed by activating the phase error and group delay error using a negative cosine function, respectively.
Finally, the phase predicted by the DNN is set as the initial phase and refined by the GLA with 100 iterations.

In the original paper \cite{takamichi2018phase,takamichi2020phase}, Takamichi \MakeLowercase{\textit{et al.}} have proven that the von Mises distribution DNN-based method significantly outperforms the plain GLA.
They also evaluate the effect of the GLA phase refinement, and the experimental results show that the refinement operation is necessary because the phase predicted by the DNN directly is unsatisfactory.

%\subsection{HiFi-GAN Vocoder-based Method}
%\label{subsec: HiFi-GAN vocoder-based Method}
%
%By simply replacing the input mel spectrograms with the amplitude spectra, the HiFi-GAN vocoder \cite{kong2020hifi} can also be regarded as a phase prediction model.
%In this particular scenario, it would be reasonable and equitable to compare it with our method.
%The HiFi-GAN cascades multiple upsampling layers and residual convolution networks to gradually upsample the input amplitude spectra to the sampling rate of the final waveform while performing non-causal residual convolutional operations.
%Hence, the phase prediction is implicitly incorporated within waveform prediction.
%The HiFi-GAN utilizes adversarial losses defined on the waveform instead of the phase spectra, ensuring the generation of high-fidelity waveforms.
%However, HiFi-GAN still has limitations in terms of generation efficiency, training efficiency and latency, due to the direct waveform prediction, adversarial training and non-causal convolutions, respectively.

\section{Proposed Method}
\label{sec: Proposed Methods}

\begin{figure*}
    \centering
    \includegraphics[width=\textwidth]{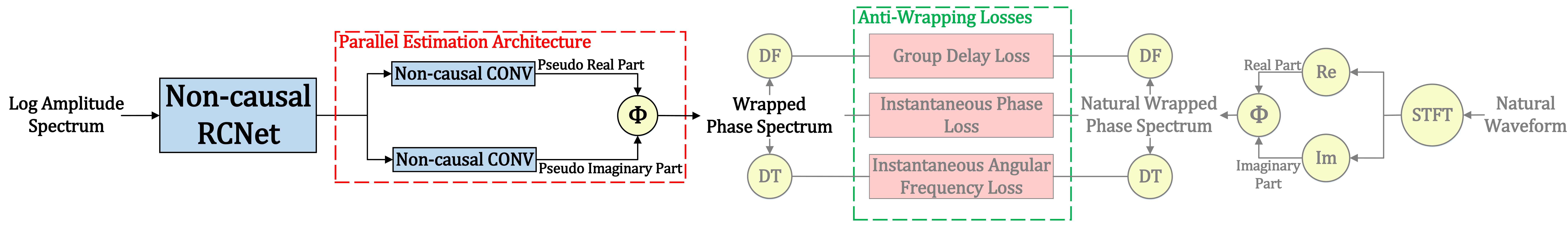}
    \caption{Details of the proposed neural speech phase prediction model. Here, \emph{RCNet}, \emph{CONV}, \emph{STFT}, \emph{DF}, \emph{DT}, \emph{Re}, \emph{Im} and \emph{$\Phi$} represent the residual convolutional network, linear convolutional layer, short-time Fourier transform, differential along frequency axis, differential along time axis, real part calculation, imaginary part calculation and phase calculation formula, respectively. Gray parts do not appear during generation.
    }
    \label{fig: Phase_model}
\end{figure*}

\begin{figure*}
    \centering
    \includegraphics[width=\textwidth]{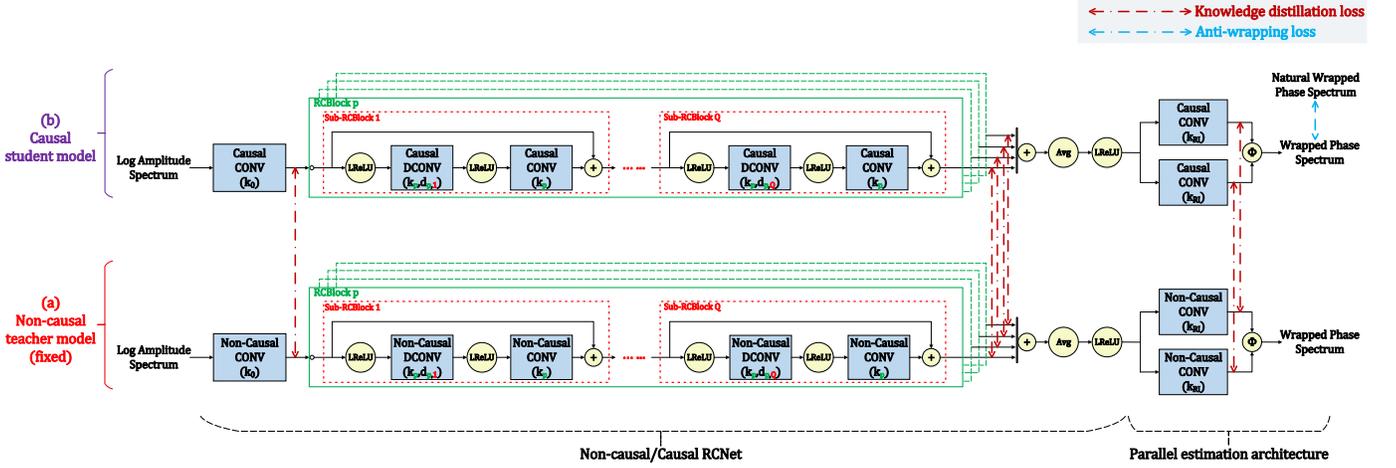}
    \caption{Details of the residual convolutional network and the training procedure of low-latency streamable neural speech phase prediction model through knowledge distillation. Here, subfigure (a) represents a non-causal teacher model which is consistent with Figure \ref{fig: Phase_model}. Subfigure (b) represents a causal student model. \emph{RCNet}, \emph{CONV}, \emph{DCONV} and \emph{$\Phi$} represent the residual convolutional network, linear convolutional layer, linear dilated convolutional layer and phase calculation formula, respectively. $k_*$ and $d_{*,*}$ denotes kernel size and dilation factor, respectively.
    }
    \label{fig: KD}
\end{figure*}

In this section, we give details on the model structure and training criteria of our proposed neural speech phase prediction model and improvements for low-latency streamable phase prediction through knowledge distillation training strategy as illustrated in Figures \ref{fig: Phase_model} and \ref{fig: KD}.

\subsection{Model Structure}
\label{subsec: Model Structure}

As shown in Figure \ref{fig: Phase_model}, the proposed neural speech phase prediction model predicts the wrapped phase spectrum $\hat{\bm{P}}\in \mathbb{R}^{F\times N}$ directly from the input log amplitude spectrum $\log\bm{A}\in \mathbb{R}^{F\times N}$ by a cascade of a non-causal residual convolutional network (RCNet) and a parallel estimation architecture.

As shown in Figure \ref{fig: KD}(a), the non-causal RCNet utilizes multiple non-causal convolutional layers to effectively broaden the receptive field, thus ensuring precise restoration of the phase.
The input log amplitude spectrum sequentially passes through a linear non-causal convolutional layer (kernel size $= k_0$ and channel size $= C$) and $P$ parallel non-causal residual convolutional blocks (RCBlocks), all of which have the same input.
Then, the outputs of these $P$ RCBlocks are summed (i.e., skip connections), averaged, and finally activated by a leaky rectified linear unit (LReLU) \cite{maas2013rectifier}.
Each RCBlock is formed by a cascade of $Q$ non-causal sub-RCBlocks.
In the $q$-th sub-RCBlock of the $p$-th RCBlock ($p=1,\dots,P$ and $q=1,\dots,Q$), the input is first activated by an LReLU, then passes through a linear non-causal dilated convolutional layer (kernel size $= k_p$, channel size $= C$ and dilation factor $= d_{p,q}$), then is activated by an LReLU again, passes through a linear non-causal convolutional layer (kernel size $= k_p$ and channel size $= C$), and finally superimposes with the input (i.e., residual connections) to obtain the output.

The parallel estimation architecture is a core module for the direct prediction of wrapped phases.
It is inspired by the process of calculating the phase spectra from the real and imaginary parts of complex spectra and consists of two parallel linear non-causal convolutional layers (kernel size $= k_{RI}$ and channel size $= N$ for both layers) and a phase calculation formula $\bm{\Phi}$.
We call the outputs of the two parallel layers as the pseudo real part $\hat{\bm{R}}\in \mathbb{R}^{F\times N}$ and pseudo imaginary part $\hat{\bm{I}}\in \mathbb{R}^{F\times N}$, respectively.
Then the wrapped phase spectrum $\hat{\bm{P}}$ is calculated by $\bm{\Phi}$ as follows:
\begin{align}
\label{equ: Phase_calculate_matrix}
\hat{\bm{P}}=\bm{\Phi}(\hat{\bm{R}},\hat{\bm{I}}).
\end{align}
Equation \ref{equ: Phase_calculate_matrix} is calculated element-wise.
For $\forall R\in \mathbb{R}$ and $I\in \mathbb{R}$, we define
\begin{align}
\label{equ: Phase calculation}
\bm{\Phi}(R,I)=\arctan\left(\dfrac{I}{R}\right)-\dfrac{\pi}{2}\cdot Sgn^*(I)\cdot\left[Sgn^*(R)-1\right],
\end{align}
and $\bm{\Phi}(0,0)=0$.
$Sgn^*$ is a symbolic function defined as:
\begin{align}
\label{equ: Symbolic_function}
Sgn^*(x)=\left\{\begin{array}{rl}1,& x\ge 0\\ -1,&x<0\end{array}\right..
\end{align}
Therefore, the range of values for the phase is $-\pi<\bm{\Phi}(R,I)\leq\pi$, meaning that the phase predicted by our model is wrapped and strictly restricted to the phase principal value interval.
Obviously, the phase value does not depend on the absolute values of the pseudo real and imaginary parts but on their relative ratios and signs.

\subsection{Training Criteria}
\label{subsec: Training Criteria}

\begin{figure}
    \centering
    \includegraphics[height=5.0cm]{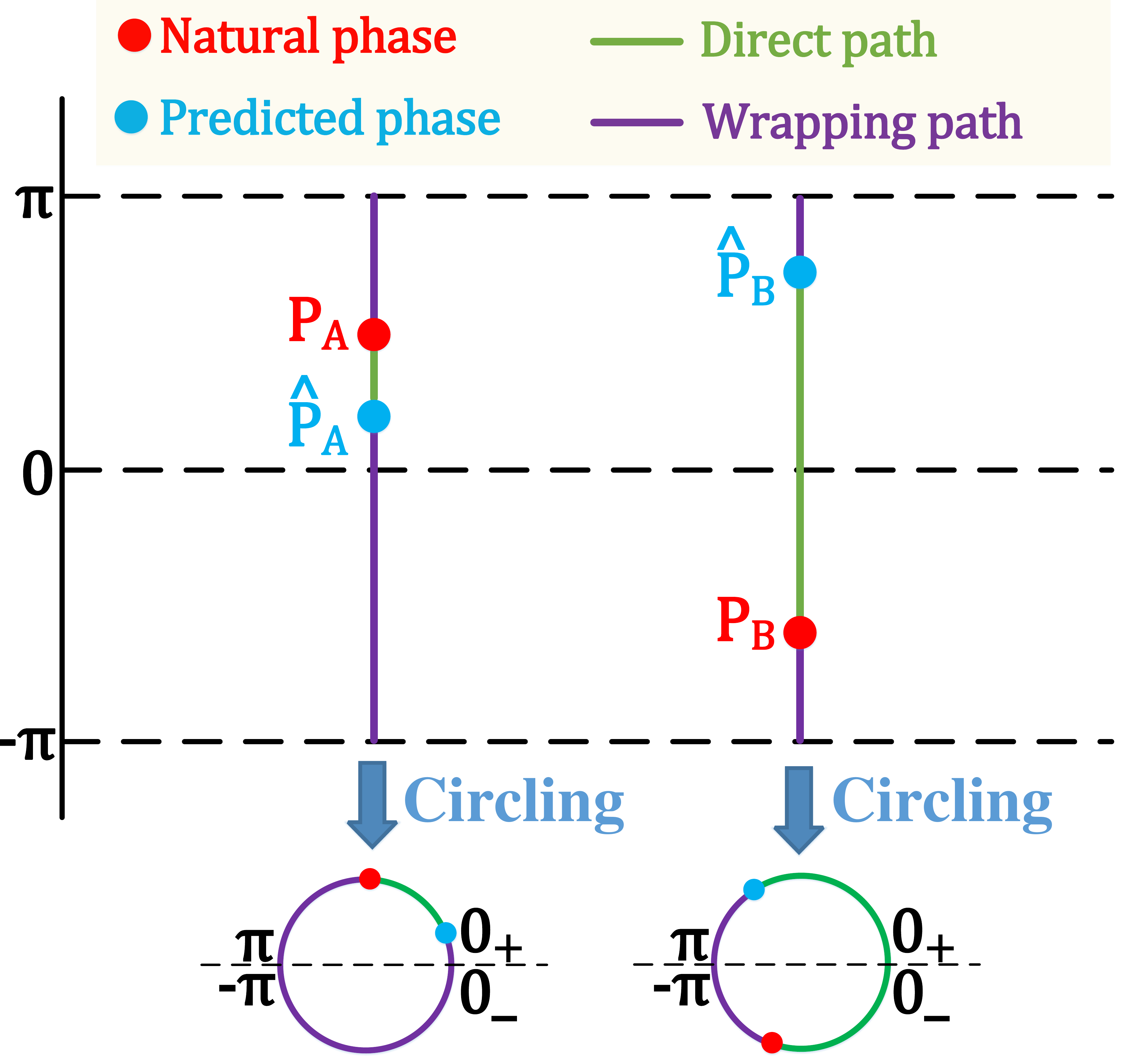}
    \caption{An illustration explanation of the error expansion issue caused by phase wrapping.
    }
    \label{fig: Phase_wrapping}
\end{figure}

Due to the wrapping property of the phase, the absolute error $e_a=|\hat{P}-P|$ between the predicted phase $\hat{P}$ and the natural phase $P$ might not be their true error.
As shown in Figure \ref{fig: Phase_wrapping}, assuming that the phase principal value interval is $(-\pi,\pi]$, there are two paths from the predicted phase point $\hat{P}_*$ to the natural one $P_*$, i.e., the direct path (corresponding to the absolute error) and the wrapping path (corresponding to the wrapping error).
Visually, we can connect the vertical line segment between $-\pi$ and $\pi$ end to end into a circle, according to the wrapping property of the phase.
Obviously, the wrapping path must pass through the boundary of the principal value interval, and the wrapping error is $e_w=2\pi-|\hat{P}-P|$.
Therefore, the true error between $\hat{P}$ and $P$ is
\begin{align}
\label{equ: True gap}
e=\min\{|\hat{P}-P|,2\pi-|\hat{P}-P|\}.
\end{align}
For example, in Figure \ref{fig: Phase_wrapping}, the true error between $\hat{P}_A$ and $P_A$ is the absolute error, but the true error between $\hat{P}_B$ and $P_B$ is the wrapping error.
This means that the absolute error and the true error satisfy $|\hat{P}-P|\ge e$, resulting in \emph{error expansion issue} when using the conventional L1 loss or mean square error (MSE) loss.
Equation \ref{equ: True gap} can be written in another form:
\begin{align}
\label{equ: True gap2}
e=\left| \hat{P}-P-2\pi\cdot round\left( \dfrac{\hat{P}-P}{2\pi} \right) \right|,
\end{align}
where $round$ represents rounding.
Obviously, Equation \ref{equ: True gap2} is a function of error $\hat{P}-P$.
We define a function $f_{line}(x)$ as follows:
\begin{align}
\label{equ: f_AW}
f_{line}(x)=\left| x-2\pi\cdot round\left( \dfrac{x}{2\pi} \right) \right|,x\in\mathbb R.
\end{align}
$f_{line}(x)$ is an anti-wrapping function which can avoid the error expansion issue caused by phase wrapping because $f_{line}(\hat{P}-P)=e$.

\begin{figure}
    \centering
    \includegraphics[height=8.3cm]{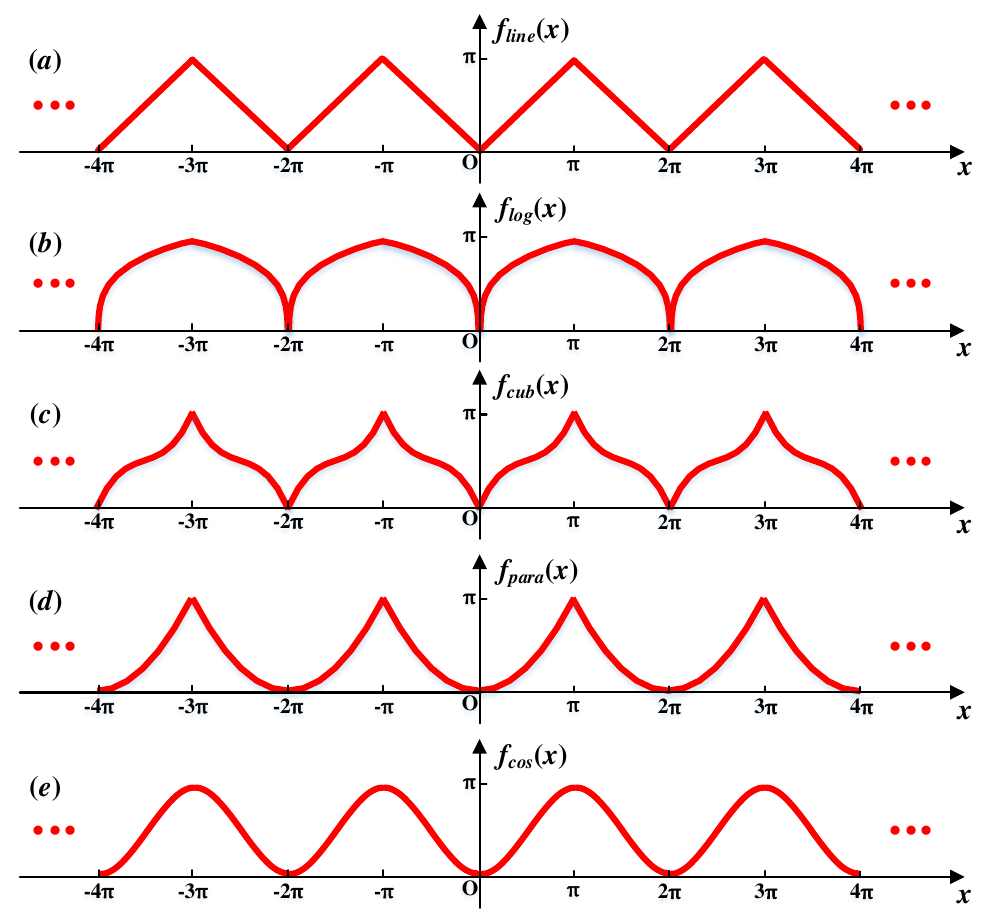}
    \caption{Graphs of five typical anti-wrapping functions, including (a) linear function; (b) logarithmic function; (c) cubic function; (d) parabolic function and (e) cosine function.
    }
    \label{fig: Anti_wrapping}
\end{figure}

As shown in Figure \ref{fig: Anti_wrapping}(a), we draw the graph of the anti-wrapping function $f_{line}(x)$.
Obviously, $f_{line}(x)$ is an even function with a period of $2\pi$ and exhibits monotonicity over half-periods.
Actually, any function $f(x)$ that satisfies below parity, periodicity and monotonicity at the same time can be used as an anti-wrapping function to activate the direct error $x$ and define loss between the predicted value and natural value.

\begin{itemize}
\item {}{\textbf{Parity}}: The anti-wrapping function $f(x)$ must be an even function because our goal is to promote the predicted value to approximate the natural value but ignore in which direction it is approximated.

\item {}{\textbf{Periodicity}}: The anti-wrapping function $f(x)$ must be a periodic function with period $2\pi$ because this periodicity cleverly avoids the problem of error expansion caused by phase wrapping.

\item {}{\textbf{Monotonicity}}: The anti-wrapping function $f(x)$ must be monotonically increasing in interval $[0,\pi]$ because the monotonicity ensures that the larger the true error $\left| x-2\pi\cdot round\left( \frac{x}{2\pi} \right) \right|$, the larger the loss $f(x)$, which conforms to the definition rules of the loss function.
\end{itemize}

Figure \ref{fig: Anti_wrapping}(b)-(e) plot several typical convex anti-wrapping functions.
%, including the logarithmic function $f_{log}(x)$, cubic function $f_{cube}(x)$, parabolic function $f_{para}(x)$ and cosine function $f_{cos}(x)$.
Compared with the linear function $f_{line}(x)$, the rate of change of a convex function may be different at different error values $x$, thereby prompting the model to pay more attention to or ignore certain ranges of error values.
In Section \ref{subsubsection: Effects of Different Anti-wrapping Functions}, we will further explore the effect of different anti-wrapping functions on model performance through experiments.

%Furthermore, our proposed anti-wrapping function is also suitable for defining the loss of the unwrapped phase.
%Suppose the predicted and natural unwrapped phases are $\hat{P}^{uw}\in \mathbb{R}$ and $P^{uw}\in \mathbb{R}$, and their corresponding wrapped phases are $\hat{P}\in (-\pi,\pi]$ and $P\in (-\pi,\pi]$, respectively.
%Then, we can obtain
%\begin{align}
%\label{equ: unwrapping_predict}
%\hat{P}^{uw}=\hat{P}+k\cdot 2\pi, k\in \mathbb{N},
%\end{align}
%\begin{align}
%\label{equ: unwrapping_natural}
%P^{uw}=P+m\cdot 2\pi, m\in \mathbb{N},
%\end{align}
%and
%\begin{align}
%\label{equ: unwrapping_difference}
%\hat{P}^{uw}-P^{uw}=\hat{P}-P+(k-m)\cdot 2\pi.
%\end{align}
%Since the anti-wrapping function $f(x)$ has a period of $2\pi$, we can derive the following formula:
%\begin{align}
%\label{equ: unwrapping_difference2}
%f(\hat{P}^{uw}-P^{uw})=f(\hat{P}-P).
%\end{align}
%Therefore, with the activation of the anti-wrapping function, the unwrapped phase loss is equivalent to the wrapped phase loss.
%The anti-wrapping function can also solve the error expansion issue caused by phase unwrapping.

Specifically, we define the instantaneous phase (IP) loss $\mathcal L_{IP}$ between the wrapped phase spectrum $\hat{\bm{P}}$ predicted by our model and the natural wrapped phase spectrum $\bm{P}=\bm{\Phi}(\bm{R},\bm{I})$ as follows:
\begin{align}
\label{equ: Phase Loss}
\mathcal L_{IP}=\mathbb{E}_{\left(\hat{\bm{P}},\bm{P}\right)} \overline{f\left(\hat{\bm{P}}-\bm{P} \right)},
\end{align}
where $f(\bm{X})$ means element-wise anti-wrapping function calculation for matrix $\bm{X}$ and $\overline{\bm{Y}}$ means averaging all elements in the matrix $\bm{Y}$.
$\bm{R}$ and $\bm{I}$ are the real and imaginary parts of the complex spectrum extracted from the natural waveform through STFT, respectively.
To ensure the continuity of the predicted wrapped phase spectrum along the frequency and time axes, we also define the group delay (GD) loss $\mathcal L_{GD}$ and instantaneous angular frequency (IAF) loss $\mathcal L_{IAF}$, which are both activated by the anti-wrapping function $f$ to avoid the error expansion issue as follows:
\begin{align}
\label{equ: Group Delay Loss}
\mathcal L_{GD}=\mathbb{E}_{\left(\Delta_{DF}\hat{\bm{P}},\Delta_{DF}\bm{P}\right)} \overline{f
\left(\Delta_{DF}\hat{\bm{P}}-\Delta_{DF}\bm{P} \right)},
\end{align}
\begin{align}
\label{equ: PTD Loss}
\mathcal L_{IAF}=\mathbb{E}_{\left(\Delta_{DT}\hat{\bm{P}},\Delta_{DT}\bm{P}\right)} \overline{f
\left(\Delta_{DT}\hat{\bm{P}}-\Delta_{DT}\bm{P} \right)},
\end{align}
where $\Delta_{DF}$ and $\Delta_{DT}$ represent the differential along the frequency axis and time axis, respectively.
Specifically, in Equation \ref{equ: Group Delay Loss}, we have
\begin{align}
\label{equ: group delay calculate}
\Delta_{DF}\hat{\bm{P}}=\hat{\bm{P}}\bm{W},\\
\Delta_{DF}\bm{P}=\bm{P}\bm{W},
\end{align}
and
\begin{align}
\label{equ: group delay matrix}
\bm{W}=\left[\bm{w}_1,\dots,\bm{w}_n,\dots,\bm{w}_{N}\right],\\
\bm{w}_n=\left[\mathop{0}_{\text{1st}},\dots,\mathop{0}_{},\mathop{1}_{n\text{-th}},\mathop{-1}_{},\mathop{0}_{},\dots,
\mathop{0}_{N\text{-th}} \right]^\top.
\end{align}
In Equation \ref{equ: PTD Loss}, we have
\begin{align}
\label{equ: PTD calculate}
\Delta_{DT}\hat{\bm{P}}=\bm{V}\hat{\bm{P}},\\
\Delta_{DT}\bm{P}=\bm{V}\bm{P},
\end{align}
and
\begin{align}
\label{equ: PTD matrix}
\bm{V}=\left[\bm{v}_1,\dots,\bm{v}_f,\dots,\bm{v}_F\right]^\top,\\
\bm{v}_f=\left[\mathop{0}_{\text{1st}},\dots,\mathop{0}_{},\mathop{1}_{f\text{-th}},\mathop{-1}_{},\mathop{0}_{},\dots,
\mathop{0}_{F\text{-th}} \right]^\top.
\end{align}
Finally, the training criteria of our proposed neural speech phase prediction model are to minimize the final loss
\begin{align}
\label{equ: Total Loss}
\mathcal L=\mathcal L_{IP}+\mathcal L_{GD}+\mathcal L_{IAF}.
\end{align}

At the generation stage, first, the well-trained neural speech phase prediction model uses the log amplitude spectrum $\log\bm{A}$ as input and predicts the wrapped phase spectrum $\hat{\bm{P}}$.
Then, the amplitude spectrum $\bm{A}$ and predicted phase spectrum $\hat{\bm{P}}$ are combined to a complex spectrum, and finally, the complex spectrum is converted to a waveform $\hat{\bm{x}}$ through ISTFT, i.e.,
\begin{align}
\label{equ: generation}
\hat{\bm{x}}=ISTFT\left(\bm{A}\odot e^{j\hat{\bm{P}}} \right).
\end{align}

\subsection{Low-Latency Streamable Phase Prediction by Causal Convolution and Knowledge Distillation}
\label{subsec: Zero-latency Phase Prediction by Knowledge Distillation}

Some application scenarios have strict requirement on the latency and streamable inference mode such as real-time voice communication.
The latency indicates the minimum amount of time needed for the model to initiate its operations.
The proposed neural speech phase prediction model incorporates non-causal convolutions to enhance its modeling capacity.
However, this inevitably results in increased latency.
For a non-causal convolution operation with a kernel size of $k$ and dilation factor of $d$, the number of future input samples required is
\begin{align}
\label{equ: future_sample}
\zeta(k,d)=\left\lfloor \dfrac{(k-1)d}{2} \right\rfloor,
\end{align}
where $\lfloor\cdot\rfloor$ denotes flooring.
Therefore, the latency measured in milliseconds of the proposed model is
\begin{align}
\label{equ: latency_NSPP}
\begin{split}
l_{NSPP}&=\{\zeta(k_0,1)+\max_{p=1,\dots,P}\left[\sum_{q=1}^Q\zeta(k_p,d_{p,q})+Q\zeta(k_p,1)  \right]
\\
&+\zeta(k_{RI},1)\}\cdot w_s,
\end{split}
\end{align}
where $w_s$ is the window shift in milliseconds of the amplitude and phase spectra.
It can be seen that when the window shift is long and the kernel size and dilation factor of convolutional layers are large, it will result in significant latency, which is undesirable in low-latency scenarios.

Therefore, as shown in Figure \ref{fig: KD}(b), we design a causal neural speech phase prediction model which can support low-latency streamable inference.
It replaces all non-causal convolutions in the non-causal model as shown in Figure \ref{fig: KD}(a) with causal convolutions.
Notably, the inference process of the causal model requires at least one frame of log amplitude spectrum input to initiate, thus, it incurs an inevitable latency equal to the window size (i.e., low latency).
However, the use of causal convolutions, which cannot leverage future information, will inevitably lead to a reduction in phase prediction precision, despite achieving low latency.
To bridge the gap between causal and non-causal models, we propose a knowledge distillation training strategy in which a non-causal teacher model guides the training of a causal student model.
Specifically, we first train a non-causal neural speech phase prediction model (i.e., the teacher model) using the anti-wrapping loss depicted in Equation \ref{equ: Total Loss}.
Then, the non-causal teacher model fixes its parameters and provide training objectives for the causal neural speech phase prediction model (i.e., the student model).
We define the output of the input convolutional layer, the outputs of $P$ RCBlocks, the pseudo real part and the pseudo imaginary part of the student model as $\hat{\bm{O}}^{I}\in \mathbb{R}^{F\times C}$, $\hat{\bm{O}}^{RCB}_{p}\in \mathbb{R}^{F\times C} (p=1,\dots,P)$, $\hat{\bm{O}}^{PRP}\in \mathbb{R}^{F\times N}$ and $\hat{\bm{O}}^{PIP}\in \mathbb{R}^{F\times N}$, respectively.
The outputs of the teacher model at corresponding positions are respectively denoted as $\tilde{\bm{O}}^{I}$, $\tilde{\bm{O}}^{RCB}_{p}$, $\tilde{\bm{O}}^{PRP}$ and $\tilde{\bm{O}}^{PIP}$.
The knowledge distillation loss is defined as follows:
\begin{align}
\label{equ: KD_loss}
\begin{split}
\mathcal L_{KD}&=\mathbb{E}_{\left(\hat{\bm{O}}^{I},\tilde{\bm{O}}^{I}\right)}\overline{\left(\hat{\bm{O}}^{I}-\tilde{\bm{O}}^{I} \right)^2}\\
&+\sum_{p=1}^P \mathbb{E}_{\left(\hat{\bm{O}}^{RCB}_{p},\tilde{\bm{O}}^{RCB}_{p}\right)}\overline{\left(\hat{\bm{O}}^{RCB}_{p}-\tilde{\bm{O}}^{RCB}_{p} \right)^2}
\\
&+\mathbb{E}_{\left(\hat{\bm{O}}^{PRP},\tilde{\bm{O}}^{PRP}\right)}\overline{\left(\hat{\bm{O}}^{PRP}-\tilde{\bm{O}}^{PRP} \right)^2}\\
&+\mathbb{E}_{\left(\hat{\bm{O}}^{PIP},\tilde{\bm{O}}^{PIP}\right)}\overline{\left(\hat{\bm{O}}^{PIP}-\tilde{\bm{O}}^{PIP} \right)^2}.
\end{split}
\end{align}
%As shown in Figure \ref{fig: KD}, we calculate the mean square error (MSE) between the output of the input convolutional layer
The training target of the student model is to minimize a combination of the anti-wrapping loss and knowledge distillation loss, i.e.,
\begin{align}
\label{equ: Total Loss_KD}
\mathcal L_{Student}=\mathcal L_{IP}+\mathcal L_{GD}+\mathcal L_{IAF}+\alpha_{KD}\mathcal L_{KD},
\end{align}
where $\alpha_{KD}$ is a hyperparameter.
Through training, the causal student model aims to approach the phase prediction capability of the non-causal teacher model while maintaining its advantage of low latency and streamable inference.

\section{Experiments}
\label{sec: Experiments}

\subsection{Data and Feature Configuration}
\label{subsec: Data and feature configuration}

A subset of the VCTK corpus \cite{veaux2016superseded} was adopted in our experiments\footnote{Source codes are available at \url{https://github.com/yangai520/LL-NSPP}. Examples of generated speech can be found at \url{https://yangai520.github.io/LL-NSPP}.}.
We selected 11,572 utterances from 28 speakers and randomly divided them into a training set (11,012 utterances) and a validation set (560 utterances).
We then built the test set, which included 824 utterances from 2 unseen speakers (a male speaker and a female speaker).
The original waveforms were downsampled to 16 kHz for the experiments.
When extracting the amplitude spectra and phase spectra from natural waveforms, the window size was 20 ms, the window shift was 5 ms (i.e., $w_s=5$), and the FFT point number was 1024 (i.e., $N=513$).

\subsection{Speech Generation Tasks}
\label{subsec: Speech Generation Tasks}

\begin{figure*}
    \centering
    \includegraphics[height=3.7cm]{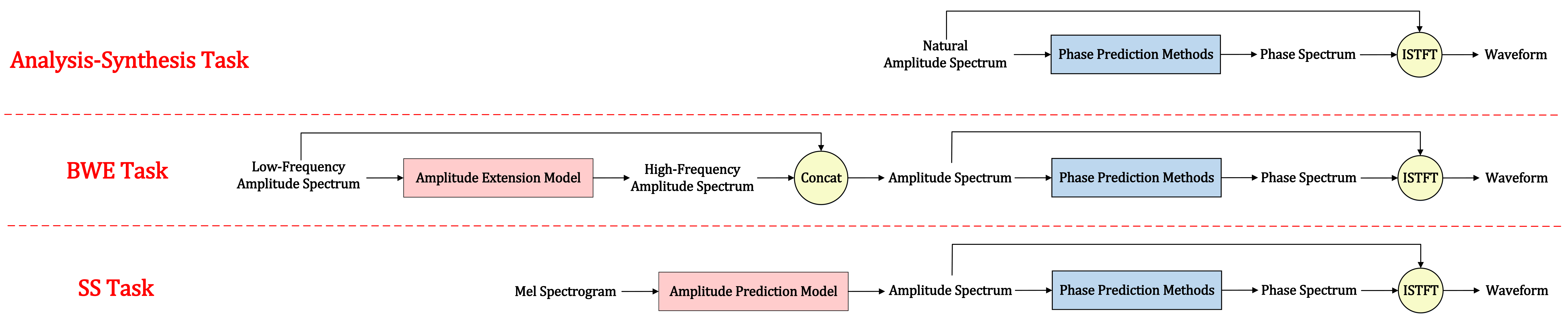}
    \caption{A simple flowchart of the analysis-synthesis task, BWE task and SS task. Here, \emph{Concat} and \emph{ISTFT} represent concatenation and inverse short-time Fourier transform, respectively.
    }
    \label{fig: Three_tasks}
\end{figure*}

In our experiments, we apply contrastive phase prediction methods to the analysis-synthesis task and two specific speech generation tasks, including the BWE task and SS task.
%three speech generation tasks, including analysis-synthesis task, BWE task and SS task.
Figure \ref{fig: Three_tasks} draws a simple flowchart of three tasks.
Specifically, the detailed description of these three tasks is as follows.

\subsubsection{Analysis-Synthesis Task}

As shown in Figure \ref{fig: Three_tasks}, the analysis-synthesis task just recovered the 513-dimensional phase spectrum from the 513-dimensional natural amplitude spectrum by phase prediction methods and reconstructed the waveform by ISTFT.

\subsubsection{BWE Task}

As shown in Figure \ref{fig: Three_tasks}, the BWE task first adopted an amplitude extension model to predict the 256-dimensional high-frequency amplitude spectrum from the 257-dimensional low-frequency amplitude spectrum.
Then, the 513-dimensional full-band amplitude spectrum was built by concatenating the low- and high-frequency amplitude spectra.
Finally, the 513-dimensional phase spectrum was recovered from the full-band amplitude spectrum by phase prediction methods and the waveform was reconstructed by ISTFT.
Here, the amplitude extension model was borrowed from our previous work \cite{ai2022denoising} and included 2 bidirectional gated recurrent unit (GRU)-based recurrent layers, each with 1024 nodes (512 forward ones and 512 backward ones), 2 convolutional layers, each with 2048 nodes (filter width=9), and a feed-forward linear output layer with 256 nodes. The generative adversarial network (GAN) with two discriminators which conducted convolution along the frequency and time axis \cite{ai2022denoising} was applied to the amplitude extension model at the training stage.

\subsubsection{SS Task}

For the SS task, we designed a neural vocoder framework with hierarchical generation of amplitude and phase spectra for statistical parametric speech synthesis (SPSS).
As shown in Figure \ref{fig: Three_tasks}, the vocoder framework first used an amplitude prediction model to complete the mapping from the 80-dimensional mel spectrogram to the 513-dimensional amplitude spectrum.
Then, the 513-dimensional phase spectrum was recovered from the amplitude spectrum by phase prediction methods and the waveform was reconstructed by ISTFT.
Here, the amplitude prediction model adopted the same structure as that used in the BWE task, except that the number of nodes in the feed-forward linear output layer was 513.

\begin{table*}
\centering
    \caption{Objective and subjective evaluation results among phase prediction methods for the analysis-synthesis task. Here, ``$a\times$" represents $a\times$ real time.}
    %\resizebox{8.8cm}{1.05cm}{
    \begin{tabular}{c | c c | c c c |  c | c}
        \hline
        \hline
        & SNR(dB)$\uparrow$ & F0-RMSE(cent)$\downarrow$ & IP loss$\downarrow$ & GD loss$\downarrow$ & IAF loss$\downarrow$ & MOS$\uparrow$ & RTF$\downarrow$\\
        \hline
        \textbf{Natural Speech} & -- & -- & -- & -- & -- & 3.93$\pm$0.063 & -- \\
        \hline
        \textbf{NSPP} & \textbf{8.26} & \textbf{10.0} & \textbf{1.479} & 0.297 & 0.694 & \textbf{3.86$\pm$0.065}& \textbf{0.051 (19.6$\times$)}\\
        \textbf{GL22} & 2.70 & 66.4 & 1.570 & 0.302 & 0.768 & 2.07$\pm$0.073 & 0.053 (18.9$\times$)\\
        \textbf{RAAR13} & 2.00 & 97.5 & 1.570 & 0.546 & 0.871 & 1.89$\pm$0.065 &  0.054 (18.5$\times$)\\
        \textbf{GL100} & 3.35 & 32.5 & 1.569 & 0.218 & 0.505 & 3.46$\pm$0.074 & 0.23 (4.48$\times$)\\
        \textbf{RAAR100} & 4.66 & 11.0 & 1.567 & \textbf{0.179} & \textbf{0.271} & \textbf{3.89$\pm$0.065} &  0.40 (2.48$\times$)\\
        \textbf{DNN+GL100} & 5.03 & 13.2 & 1.537 & 0.209 & 0.484 & 3.70$\pm$0.068 & 0.29 (3.45$\times$)\\
        \hline
        \hline
    \end{tabular}%}
\label{tab_results_AS}
\end{table*}

\subsection{Comparison among Phase Prediction Methods}
\label{subsec: Comparison among Phase Prediction Methods}

We conducted objective and subjective experiments to compare the performance of our proposed neural speech phase prediction model and other phase prediction methods for the analysis-synthesis task, BWE task and SS task.
The descriptions of the phase prediction methods for comparison are as follows:

\begin{itemize}
\item {}{\textbf{NSPP}}: The proposed neural speech phase prediction model with latency as shown in Figure \ref{fig: Phase_model}.
In the non-causal RCNet, the kernel size of the input linear convolutional layer was $k_0=7$.
There were 3 parallel RCBlocks (i.e., $P=3$) in the RCNet, and each RCBlock was formed by concatenating 3 sub-RCBlocks (i.e., $Q=3$).
The kernel sizes of RCBlocks were $k_1=3$, $k_2=7$ and $k_3=11$, and the dilation factors of sub-RCBlocks within each RCBlock were $d_{*,1}=1$, $d_{*,2}=3$ and $d_{*,3}=5$.
The channel size of all the convolutional operations in the RCNet was $C=512$.
In the parallel estimation architecture, the kernel size of two parallel linear convolutional layers was $k_{RI}=7$.
We used the linear anti-wrapping function $f_{line}(x)$ as shown in Figure \ref{fig: Anti_wrapping}(a) at the training stage.
$f_{line}(x)$ was given in Equation \ref{equ: f_AW}.
The model was trained using the AdamW optimizer \cite{loshchilov2018decoupled} with $\beta_1=0.8$ and $\beta_2=0.99$ on a single Nvidia 3090Ti GPU until 3100 epochs.
The learning rate decay was scheduled by a 0.999 factor in every epoch with an initial learning rate of 0.0002.
The batch size was 16, and the truncated waveform length was 8000 samples (i.e., 0.5 s) for each training step.
Based on the current configuration, the \textbf{NSPP} exhibited a latency of 330 ms, as calculated using Equation \ref{equ: latency_NSPP}.

\item {}{\textbf{GL$\bm{n}$}}: The GLA \cite{griffin1984signal} mentioned in Section \ref{subsec: Griffin-Lim algorithm (GLA)} with $n$ iterations ($n=22$ and $n=100$ were used in the experiments). The GLA required the amplitude spectra of an entire utterance as input, thus the latency of the \textbf{GL$\bm{n}$} equaled to utterance length $T$ in milliseconds.

\item {}{\textbf{RAAR$\bm{n}$}}: The RAAR \cite{kobayashi2022acoustic} mentioned in Section \ref{subsec: Relaxed averaged alternating reflection (RAAR)} with $n$ iterations ($n=13$ and $n=100$ were used in the experiments).
    Same as the \textbf{GL$\bm{n}$}, the latency of the \textbf{RAAR$\bm{n}$} was also equal to $T$.

\item {}{\textbf{DNN+GL100}}: The von Mises distribution DNN-based phase prediction method \cite{takamichi2018phase,takamichi2020phase} mentioned in Section \ref{subsec: Von-Mises-distribution DNN-based method}.
    The phase spectra were first predicted by the DNN and then refined by the GLA with 100 iterations.
    We reimplemented it ourselves.
    The training configuration of the DNN is the same as that of \textbf{NSPP}.
    %The latency of the \textbf{DNN+GL100} corresponded to the maximum value between the latencies of the DNN and GLA.
    As mentioned in Section \ref{subsec: Von-Mises-distribution DNN-based method}, the DNN adopted the amplitude spectra at current and $\pm$2 frames, resulting in the latency of 2$w_s$=10 ms.
    The latency of the \textbf{DNN+GL100} corresponded to the maximum value between the latencies of the DNN and GLA, i.e., $max\{10, T \}$.

\end{itemize}

\begin{table}
\centering
    \caption{Objective and subjective evaluation results among phase prediction methods for the BWE task.}
    %\resizebox{8.8cm}{1.05cm}{
    \begin{tabular}{c | c c | c}
        \hline
        \hline
        & SNR(dB)$\uparrow$ & F0-RMSE(cent)$\downarrow$ & MOS$\uparrow$\\
        \hline
        \textbf{Natural Speech} & -- & -- & 4.15$\pm$0.050\\
        \hline
        \textbf{NSPP} & \textbf{8.18} & \textbf{10.8} & \textbf{4.09$\pm$0.052}\\
        \textbf{GL100} & 3.24 & 32.6 & 3.90$\pm$0.069\\
        \textbf{RAAR100} & 4.49 & 11.0 & \textbf{4.10$\pm$0.053}\\
        \textbf{DNN+GL100} & 5.03 & 13.2 & 4.02$\pm$0.059\\
        \hline
        \hline
    \end{tabular}%}
\label{tab_results_BWE}
\end{table}

To objectively evaluate the phase prediction precision, we calculated the average IP, GD and IAF losses on the test set.
To objectively evaluate the reconstructed speech quality, two objective metrics used in our previous work \cite{ai2020neural} were adopted here, including the signal-to-noise ratio (SNR), which was an overall measurement of the distortions of both amplitude and phase spectra, and root MSE of F0 (denoted by F0-RMSE), which reflected the distortion of F0.
To evaluate the generation efficiency, the real-time factor (RTF), which is defined as the ratio between the time consumed to generate all test sentences using a single Intel Xeon E5-2680 CPU core and the total duration of the test set, was also utilized as an objective metric.
Regarding the subjective evaluation, Mean opinion score (MOS) tests were conducted to compare the naturalness of the speeches reconstructed by these methods.
In each MOS test, twenty test utterances reconstructed by these methods along with the natural utterances were evaluated by at least 30 native English listeners on the crowdsourcing platform of Amazon Mechanical Turk\footnote{\url{https://www.mturk.com}.} with anti-cheating considerations \cite{buchholz2011crowdsourcing}.
Listeners were asked to give a naturalness score between 1 and 5, and the score interval was 0.5.

\begin{table}
\centering
    \caption{Objective and subjective evaluation results among phase prediction methods for the SS task.}
    %\resizebox{8.8cm}{1.05cm}{
    \begin{tabular}{c | c c | c}
        \hline
        \hline
        & SNR(dB)$\uparrow$ & F0-RMSE(cent)$\downarrow$ & MOS$\uparrow$\\
        \hline
        \textbf{Natural Speech} & -- & -- & 3.84$\pm$0.051\\
        \hline
        \textbf{NSPP} & \textbf{6.75} & \textbf{19.0} & \textbf{3.73$\pm$0.055}\\
        \textbf{GL100} & 3.14 & 39.4 & 3.50$\pm$0.068\\
        \textbf{RAAR100} & 3.92 & 22.7 & 3.64$\pm$0.061\\
        \textbf{DNN+GL100} & 4.02 & 22.5 & 3.66$\pm$0.062\\
        \hline
        \hline
    \end{tabular}%}
\label{tab_results_SS}
\end{table}

For the analysis-synthesis task, BWE task and SS task, both the objective and subjective results are listed in Table \ref{tab_results_AS}, Table \ref{tab_results_BWE} and Table \ref{tab_results_SS}, respectively.
Our proposed \textbf{NSPP} obtained the highest SNR and the lowest F0-RMSE among all methods for all three tasks.
The IP loss, GD loss and IAF loss are only calculated for the analysis-synthesis task.
Our proposed \textbf{NSPP} obtained the lowest IP loss but felled behind in two other metrics when compared to iterative algorithms.
This indicates that our proposed model primarily achieved precise phase prediction by improving the IP loss compared with other methods.
In our experiments, we discovered that reducing IP loss is challenging, which can be attributed to the sensitivity of instantaneous phase to waveform shifts \cite{masuyama2020phase}.
Regarding the RTF results shown in Table \ref{tab_results_AS}, our proposed \textbf{NSPP} was also an efficient model, reaching 19.6x real-time generation on a CPU.
At the same generation speed, the GLA and RAAR could only iterate 22 rounds and 13 rounds (i.e., \textbf{GL22} and \textbf{RAAR13}), respectively, and their reconstructed speech quality was far inferior to that of \textbf{NSPP}.
It is also worth mentioning that the training speed of the \textbf{NSPP} was also fast, with a training time of 27 hours on this dataset using a single Nvidia 3090Ti GPU.
Regarding the subjective results, the MOS score of the \textbf{NSPP} approached that of the natural speech for the analysis-synthesis task as shown in Table \ref{tab_results_AS}, and the difference between the \textbf{NSPP} and \textbf{Natural Speech} was slightly insignificant ($p=0.055$ of paired $t$-tests).
The \textbf{GL100}, although fully iterated, still performed significantly worse than our proposed \textbf{NSPP} ($p<0.01$) for all three tasks due to the audible unnatural artifact sounds.
Compared with the \textbf{GL100}, the performance of the \textbf{DNN+GL100} was significantly improved ($p<0.01$), which was consistent with the conclusion in the original paper \cite{takamichi2018phase,takamichi2020phase}.
Nevertheless, our proposed \textbf{NSPP} still outperformed \textbf{DNN+GL100} in terms of both the reconstructed speech quality and generation speed for all three tasks.
These results proved the precise phase prediction ability of our proposed model.
Besides, compared with the \textbf{DNN+GL100}, the proposed \textbf{NSPP} was a fully neural network-based method without the extra phase refinement operation, which can be easily implemented.
However, the subjective differences between the \textbf{NSPP} and \textbf{RAAR100} were not significant for both the analysis-synthesis task ($p=0.38$) and the BWE task ($p=0.98$).
Interestingly, for the SS task, the MOS score of the \textbf{NSPP} was significantly higher than that of the \textbf{RAAR100} ($p<0.01$).
Obviously, the amplitude spectra used to recover the phase spectra in the analysis-synthesis task and BWE task were natural and semi-natural, respectively, but the amplitude spectra in the SS task were completely degraded.
These results illustrated that our proposed \textbf{NSPP} had good robustness, while the quality of the phase spectra recovered by the iterative algorithms (i.e., the GLA and RAAR) from the degraded amplitude spectra were obviously restricted.
The proposed neural speech phase prediction model was more suitable for specific speech generation tasks.

\subsection{Comparison with Waveform Reconstruction Method}
\label{subsec: Comparison with Waveform Reconstruction Method}

Unlike iterative and neural network-based phase prediction methods, the waveform reconstruction methods were not originally designed for phase prediction.
However, these waveform reconstruction methods implicitly incorporated phase prediction within waveform prediction.
In this subsection, we compared our proposed \textbf{NSPP} with the HiFi-GAN vocoder (denoted by \textbf{HiFi-GAN}) using both objective and subjective evaluations.
The description of the \textbf{HiFi-GAN} is as follows:

\begin{itemize}
%\item {}{\textbf{NSPP}}: The proposed neural speech phase prediction model.
     %Model structures and settings were the same as \textbf{NSPP-line} in Section \ref{subsec: Effects of different anti-wrapping functions}.

\item {}{\textbf{HiFi-GAN}}: The v1 version of the HiFi-GAN vocoder \cite{kong2020hifi}.
We reimplemented it using the open source implementation\footnote{\url{https://github.com/jik876/hifi-gan}.}.
We made small modifications to the open source code to fit our configurations.
For a fair comparison with the \textbf{NSPP}, the input of the \textbf{HiFi-GAN} is 513-dimensional log amplitude spectra rather than 80-dimensional mel spectrograms.
The upsampling ratios were set as $h_1=5$, $h_2=4$, $h_3=2$ and $h_4=2$.
The latency calculation manner for the \textbf{HiFi-GAN} is similar with the \textbf{NSPP}.
Although the \textbf{HiFi-GAN} incorporated more convolutional layers, the majority of its operations are conducted at a higher sampling rate (i.e., $w_s$ is much smaller in Equation \ref{equ: latency_NSPP}) relative to the original amplitude spectrum. Consequently, the latency of the \textbf{HiFi-GAN} amounted to a mere 101.4375 ms.%, which is significantly higher than that of the \textbf{NSPP}.

\end{itemize}

The objective evaluation results for the analysis-synthesis task are listed in Table \ref{tab_results_HiFi-GAN}.
Our proposed \textbf{NSPP} slightly outperformed \textbf{HiFi-GAN} on the SNR and F0-RMSE metrics.
However, considering the phase prediction precision, the phase continuity of the proposed \textbf{NSPP} was significantly superior to that of the \textbf{HiFi-GAN} according to the results of the GD loss and IAF loss, which confirmed the effectiveness of our proposed direct phase prediction manner.
Regarding the generation efficiency, the RTF of \textbf{HiFi-GAN} on GPU was comparable to our proposed \textbf{NSPP} because GPUs allowed for parallel accelerated computations.
However, on CPU, the \textbf{NSPP} exhibited significantly higher generation efficiency compared to the \textbf{HiFi-GAN}.
Besides, due to the absence of GAN, the training time of \textbf{NSPP} is also much shorter than that of \textbf{HiFi-GAN} when using the same training mode and total number of training epochs (i.e., 3100 epochs).
This validated the efficiency advantage of the \textbf{NSPP}.

\begin{figure}
    \centering
    \includegraphics[height=3.5cm]{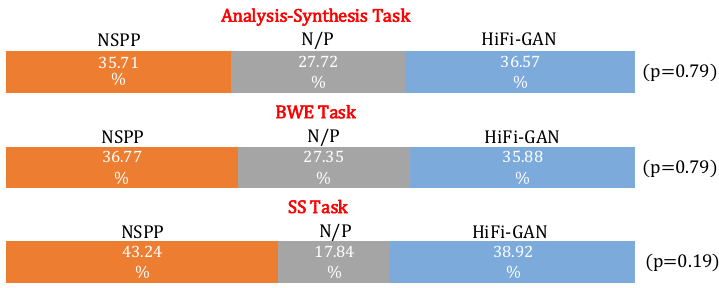}
    \caption{Average preference scores (\%) of ABX tests on speech quality between \textbf{NSPP} and \textbf{HiFi-GAN}, where N/P stands for ``no preference" and $p$ denotes the $p$-value of a $t$-test between two models.
    }
    \label{fig: ABX_HiFiGAN}
\end{figure}

Regarding the subjective evaluations, we conducted ABX preference tests on the Amazon Mechanical Turk platform to compare the subjective quality of the speeches generated by the \textbf{NSPP} and \textbf{HiFi-GAN}.
In each ABX test, twenty utterances were randomly selected from the test set reconstructed by two comparative models and evaluated by at least 30 native English listeners.
The listeners were asked to judge which utterance in each pair had better speech quality or whether there was no preference.
In addition to calculating the average preference scores, the $p$-value of a $t$-test was used to measure the significance of the difference between two models.
The results are shown in Figure \ref{fig: ABX_HiFiGAN}.
There was no significant difference ($p > 0.01$) in subjective perception between the \textbf{NSPP} and \textbf{HiFi-GAN}, whether in analysis-synthesis, BWE or SS tasks.
This finding suggests that the \textbf{NSPP} was on par with the \textbf{HiFi-GAN} in terms of reconstructed speech quality and robustness, while also offering a remarkable efficiency advantage.

\begin{table}
\centering
    \caption{Objective evaluation results between \textbf{NSPP} and \textbf{HiFi-GAN} for the analysis-synthesis task. Here, ``$a\times$" represents $a\times$ real time.}
    %\resizebox{8.8cm}{1.05cm}{
    \begin{tabular}{c | c c }
        \hline
        \hline
        & \textbf{NSPP} & \textbf{HiFi-GAN} \\
        \hline
        SNR(dB)$\uparrow$ & \textbf{8.26} & 7.37\\
        F0-RMSE(cent)$\downarrow$ & \textbf{10.0} & 13.2 \\
        \hline
        IP loss$\downarrow$ & \textbf{1.479} & 1.483 \\
        GD loss$\downarrow$ & \textbf{0.297} &  0.352\\
        IAF loss$\downarrow$ & \textbf{0.694} &  1.011\\
        \hline
        RTF (GPU)$\downarrow$ & \textbf{0.0065 (154$\times$)} & 0.0092 (109$\times$)\\
        RTF (CPU)$\downarrow$ & \textbf{0.051 (19.6$\times$)} & 0.60 (1.66$\times$)\\
        \hline
        Training Time(h)$\downarrow$ & \textbf{27} & 326\\
        \hline
        \hline
    \end{tabular}%}
\label{tab_results_HiFi-GAN}
\end{table}

It should be noted that the objective of this study is not to compare with other end-to-end speech generation methods.
We are solely comparing the performance of different phase prediction methods when given different amplitude inputs.
However, due to the trainable nature of the proposed neural speech phase prediction model, it can be easily integrated into end-to-end speech generation tasks to improve the phase quality, where the APNet vocoder \cite{ai2023apnet} and MP-SENet \cite{lu2023mp} speech enhancement model serve as illustrative examples.
%For example, we have proposed the APNet vocoder \cite{ai2023apnet} which achieved parallel prediction of amplitude and phase spectra from input mel spectrograms.
%The APNet vocoder achieved an approximately 8x faster generation speed than HiFi-GAN v1 on a CPU, while its synthesized speech quality was comparable to HiFi-GAN v1.
%We have also proposed the MP-SENet \cite{lu2023mp}, a speech enhancement network which directly denoises amplitude and phase spectra in parallel, achieving a state-of-the-art perceptual evaluation of speech quality (PESQ) of 3.50 on the public VoiceBank+DEMAND dataset.

\subsection{Evaluation on Low-Latency Streamable Phase Prediction}
\label{subsec: Experimental Validation of Zero-Latency Phase Prediction}

%Based on the current configuration, the proposed neural speech phase prediction model exhibited a latency of 330 ms, as calculated using Equation \ref{equ: latency_NSPP}.
%The latency calculation manner for the HiFi-GAN vocoder is similar.
%Although the HiFi-GAN vocoder incorporates more convolutional layers, the majority of its operations are conducted at a higher sampling rate (i.e., $w_s$ is much smaller in Equation \ref{equ: latency_NSPP}) relative to the original amplitude spectrum. Consequently, the latency of the HiFi-GAN vocoder amounts to a mere 101.4375 ms.
As discussed in Section \ref{subsec: Comparison among Phase Prediction Methods} and \ref{subsec: Comparison with Waveform Reconstruction Method}, our proposed \textbf{NSPP} exhibited a distinct advantage in latency compared to the \textbf{GL$\bm{n}$}, \textbf{RAAR$\bm{n}$} and \textbf{DNN+GL100} when dealing with lengthy utterances.
However, compared to the \textbf{HiFi-GAN}, the latency of our proposed \textbf{NSPP} was somewhat disappointing.
Therefore, it is highly necessary to further reduce the latency of the proposed model, as discussed in Section \ref{subsec: Zero-latency Phase Prediction by Knowledge Distillation}.

To validate the effectiveness of the low-latency streamable phase prediction method proposed in Section \ref{subsec: Zero-latency Phase Prediction by Knowledge Distillation}, we compared the \textbf{NSPP} with the following two models:

\begin{itemize}

\item {}{\textbf{NSPP\_causal}}: The causal neural speech phase prediction model trained only using the anti-wrapping losses (i.e., Equation \ref{equ: Total Loss}).
\item {}{\textbf{NSPP\_causal\_KD}}: The causal neural speech phase prediction model trained using the combination of anti-wrapping losses and knowledge distillation losses (i.e., Equation \ref{equ: Total Loss_KD}).

\end{itemize}

\begin{figure}
    \centering
    \includegraphics[height=6cm]{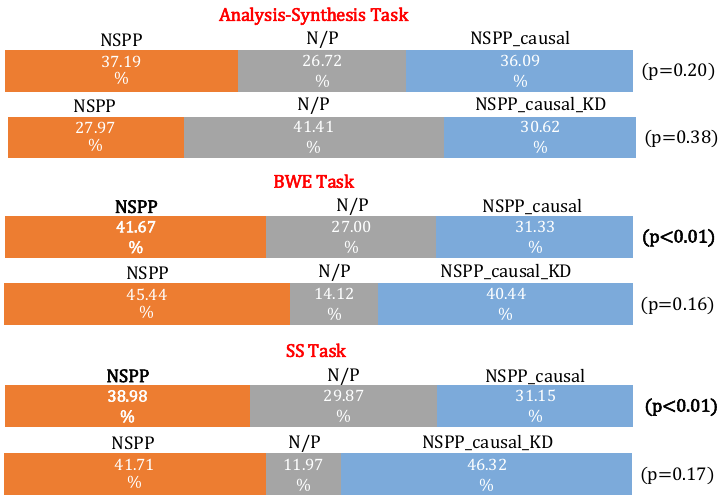}
    \caption{Average preference scores (\%) of ABX tests on speech quality for \textbf{NSPP\_causal}, \textbf{NSPP\_causal\_KD} and \textbf{NSPP}, where N/P stands for ``no preference" and $p$ denotes the $p$-value of a $t$-test between two models.
    }
    \label{fig: ABX_latency}
\end{figure}

\begin{figure}
    \centering
    \includegraphics[height=6cm]{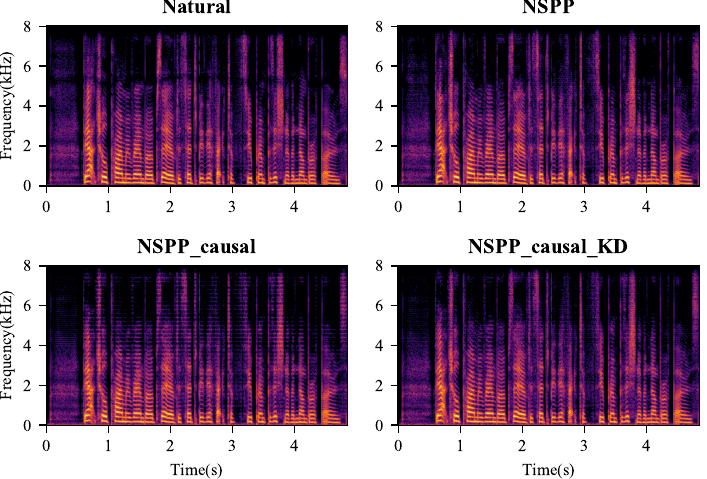}
    \caption{A comparison among the spectrograms of the natural speech and speeches generated by \textbf{NSPP}, \textbf{NSPP\_causal} and \textbf{NSPP\_causal\_KD} for the BWE task.}
    \label{fig: Spectrogram_latency}
\end{figure}

The aforementioned two models both have a 20 ms latency (i.e., the window size) and support streamable inference.
We first compared the \textbf{NSPP} and \textbf{NSPP\_causal}.
The objective (i.e., three phase losses) and subjective (i.e., ABX tests) evaluation results are shown in Table \ref{tab_results_latency} and Figure \ref{fig: ABX_latency}, respectively.
Unsurprisingly, replacing non-causal convolutions with causal convolutions led to a significant decrease in the performance of the proposed model.
Specifically, there is a noticeable increase in the GD and IAF losses of phase spectra predicted by \textbf{NSPP\_causal}.
In terms of perceptual quality, the \textbf{NSPP\_causal} lagged significantly ($p < 0.01$) behind the \textbf{NSPP} in BWE and SS tasks, indicating a lack of robustness.
To provide further evidence, we plotted the spectrograms of the reconstructed speech from the \textbf{NSPP} and \textbf{NSPP\_causal} for the BWE task.
As shown in Figure \ref{fig: Spectrogram_latency}, the \textbf{NSPP\_causal} experienced severe spectral interference, which may be the reason for the decline in auditory perception.

\begin{table}
\centering
    \caption{Objective evaluation results of \textbf{NSPP}, \textbf{NSPP\_causal} and \textbf{NSPP\_causal\_KD} for the analysis-synthesis task.}
    %\resizebox{8.8cm}{1.05cm}{
    \begin{tabular}{c | c c  c}
        \hline
        \hline
        & IP loss$\downarrow$ & GD loss$\downarrow$ & IAF loss$\downarrow$\\
        \hline
        \textbf{NSPP} & \textbf{1.479} & \textbf{0.297} & \textbf{0.694}\\
        \textbf{NSPP\_causal} & 1.497 & 0.372 & 1.02\\
        \textbf{NSPP\_causal\_KD} & 1.499 & 0.309 & 0.766\\
        \hline
        \hline
    \end{tabular}%}
\label{tab_results_latency}
\end{table}

When the causal neural speech phase prediction model is integrated into the training process with knowledge distillation loss, remarkable improvements are observed.
The GD and IAF losses of the \textbf{NSPP\_causal\_KD} approached the upper bound \textbf{NSPP} as listed in Table \ref{tab_results_latency}.
There are no significant subjective difference ($p > 0.05$) between the \textbf{NSPP} and \textbf{NSPP\_causal\_KD} for all tasks as shown in Figure \ref{fig: ABX_latency}.
Compared with the \textbf{NSPP\_causal}, the issue of spectral distortion also disappeared in the \textbf{NSPP\_causal\_KD} as shown in Figure \ref{fig: Spectrogram_latency}.
Through knowledge distillation, the student model successfully learned the knowledge from the teacher model.
%The above results strongly demonstrate that the proposed knowledge distillation strategy ensured zero-latency generation while maintaining phase prediction precision, efficiency and robustness of the proposed neural speech phase prediction model.
The above results strongly demonstrate that the combination of causal convolution and knowledge distillation reduced the latency of the proposed neural phase prediction model from a very high 330 ms to an extremely low 20 ms, while maintaining phase prediction precision, efficiency and robustness of the model.

\subsection{Discussions}
\label{subsec: Discussions}

\subsubsection{Effects of Different Anti-Wrapping Functions}
\label{subsubsection: Effects of Different Anti-wrapping Functions}

As introduced in Section \ref{subsec: Training Criteria}, any function that conforms to the properties of parity, periodicity, and monotonicity can be used as an anti-wrapping function to activate the error between the predicted value and the natural value at the training stage.
%Figure \ref{fig: Anti_wrapping} gives five typical anti-wrapping functions.
In this experiment, we studied the effect of different types of anti-wrapping functions on the performance of our proposed neural speech phase prediction model.
The models used for comparison with the \textbf{NSPP} are shown as follows.

\begin{itemize}

\item {}{\textbf{NSPP-log}}: The proposed neural speech phase prediction model using the logarithmic anti-wrapping function $f_{log}(x)$ as shown in Figure \ref{fig: Anti_wrapping}(b) at the training stage. In the primary period, $f_{log}(x)=\frac{\pi}{ln(\pi+1)}ln(x+1),x\in (-\pi,\pi]$.

\item {}{\textbf{NSPP-cub}}: The proposed neural speech phase prediction model using the cubic anti-wrapping function $f_{cub}(x)$ as shown in Figure \ref{fig: Anti_wrapping}(c) at the training stage. In the primary period, $f_{cub}(x)=\frac{4}{\pi^2}\left(x-\frac{\pi}{2} \right)^3+\frac{\pi}{2},x\in (-\pi,\pi]$.

\item {}{\textbf{NSPP-para}}: The proposed neural speech phase prediction model using the parabolic anti-wrapping function $f_{para}(x)$ as shown in Figure \ref{fig: Anti_wrapping}(d) at the training stage. In the primary period, $f_{para}(x)=\frac{1}{\pi}x^2,x\in (-\pi,\pi]$.

\item {}{\textbf{NSPP-cos}}: The proposed neural speech phase prediction model using the cosine anti-wrapping function $f_{cos}(x)$ as shown in Figure \ref{fig: Anti_wrapping}(e) at the training stage. In the primary period, $f_{cos}(x)=-\frac{\pi}{2}cos(x)+\frac{\pi}{2},x\in (-\pi,\pi]$.

\end{itemize}

The above four models and the \textbf{NSPP} shared the same settings except for the anti-wrapping function used during training.
We compared the performance of these five models using the subjective evaluation for the analysis-synthesis task.
MOS tests were conducted to compare the naturalness of the speeches reconstructed by these models.
The subjective results are listed in Table \ref{tab_results_function}.
Obviously, the \textbf{NSPP}, \textbf{NSPP-log} and \textbf{NSPP-cub} all achieved excellent performance because their MOS scores were close to the natural one.
However, \textbf{NSPP-para} and \textbf{NSPP-cos} were inferior to the other models.
As shown in Figure \ref{fig: Anti_wrapping}, the commonality of $f_{line}(x)$, $f_{log}(x)$ and $f_{cub}(x)$ is that when the true error $\left| x-2\pi\cdot round\left( \frac{x}{2\pi} \right) \right|$ is smaller, the rate of change of these functions is faster or remains the same.
The above conclusion is opposite for functions $f_{para}(x)$ and $f_{cos}(x)$.
These results indicated that an anti-wrapping function that paid less attention to activations for small error segments leads to a decrease in the phase prediction performance of the model (i.e., the $f_{para}(x)$ and $f_{cos}(x)$).
At the small error segment (e.g., $x\in \left[-\frac{\pi}{2},\frac{\pi}{2}\right]$), the rate of change of the function should be faster than the rate of change of the error.
For example, by comparing \textbf{NSPP-log} and \textbf{NSPP-para}, when they reduced the same loss value at the training stage, \textbf{NSPP-log} shrank the true error faster than \textbf{NSPP-para}.
Quickly reducing the true error is the goal of model training.
However, although the rates of change of function $f_{line}(x)$, $f_{log}(x)$ and $f_{cub}(x)$ were significantly different at the large error range (e.g., $x\in \left(-\pi,-\frac{\pi}{2}\right]\cup \left[\frac{\pi}{2},\pi\right]$), the results of \textbf{NSPP}, \textbf{NSPP-log} and \textbf{NSPP-cub} were not significantly different.
It is reasonable because we find that the true error was mostly concentrated in the small value segment (i.e., most $x\in \left[-\frac{\pi}{2},\frac{\pi}{2}\right]$).
Taking the \textbf{NSPP} as an example, the converged mean values of the true errors of the IP, GD and IAF on the test set were 1.48, 0.297 and 0.694 respectively.

\begin{table}
\centering
    \caption{Subjective evaluation results among five neural speech phase prediction models for the comparison of anti-wrapping functions on the analysis-synthesis task.}
    %\resizebox{8.8cm}{1.05cm}{
    \begin{tabular}{c | c}
        \hline
        \hline
         & MOS$\uparrow$\\
        \hline
        \textbf{Natural Speech} & 3.96$\pm$0.048 \\
        \hline
        \textbf{NSPP} & \textbf{3.91$\pm$0.049}\\
        \textbf{NSPP-log} & \textbf{3.90$\pm$0.048}\\
        \textbf{NSPP-cub} & \textbf{3.93$\pm$0.048}\\
        \textbf{NSPP-para} & 3.79$\pm$0.050\\
        \textbf{NSPP-cos} & 3.84$\pm$0.053\\
        \hline
        \hline
    \end{tabular}%}
\label{tab_results_function}
\end{table}

\subsubsection{Ablation Studies}
\label{subsubsection: Ablation Studies}

We then conducted several ablation experiments to explore the roles of some key modules in our proposed \textbf{NSPP}.
Here, experiments were performed only on the analysis-synthesis task.
The ablated variants of the \textbf{NSPP} for comparison included the following:
\begin{itemize}
\item {}{\textbf{NSPP wo PEA}}: Removing the parallel estimation architecture from the \textbf{NSPP}.
The output of the residual convolutional network passes through a linear layer without activation to predict the phase spectra, which is the same way as used in the von Mises distribution DNN-based method \cite{takamichi2018phase,takamichi2020phase}.

\item {}{\textbf{NSPP wo AWF}}: Removing the anti-wrapping function $f$ from the \textbf{NSPP} and adopting L1 losses for $\mathcal L_{IP}$, $\mathcal L_{GD}$ and $\mathcal L_{IAF}$ at the training stage.

\item {}{\textbf{NSPP wo IP}}: Removing the IP loss $\mathcal L_{IP}$ from the \textbf{NSPP} at the training stage.

\item {}{\textbf{NSPP wo GD}}: Removing the GD loss $\mathcal L_{GD}$ from the \textbf{NSPP} at the training stage.

\item {}{\textbf{NSPP wo IAF}}: Removing the IAF loss $\mathcal L_{IAF}$ from the \textbf{NSPP} at the training stage.

\end{itemize}

\begin{figure}
    \centering
    \includegraphics[height=4.0cm]{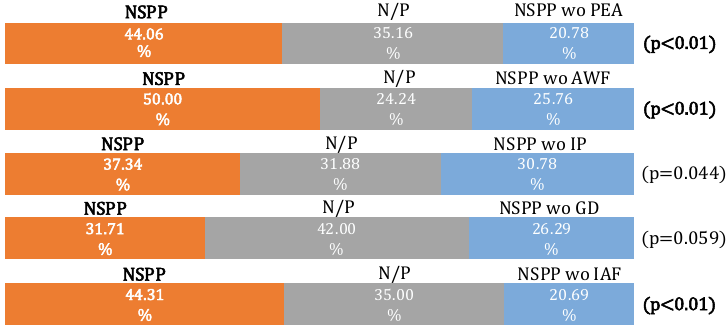}
    \caption{Average preference scores (\%) of ABX tests on speech quality between \textbf{NSPP} and its ablated variants for the analysis-synthesis task, where N/P stands for ``no preference" and $p$ denotes the $p$-value of a $t$-test between two models.
    }
    \label{fig: ABX}
\end{figure}

\begin{figure}
    \centering
    \includegraphics[height=\linewidth]{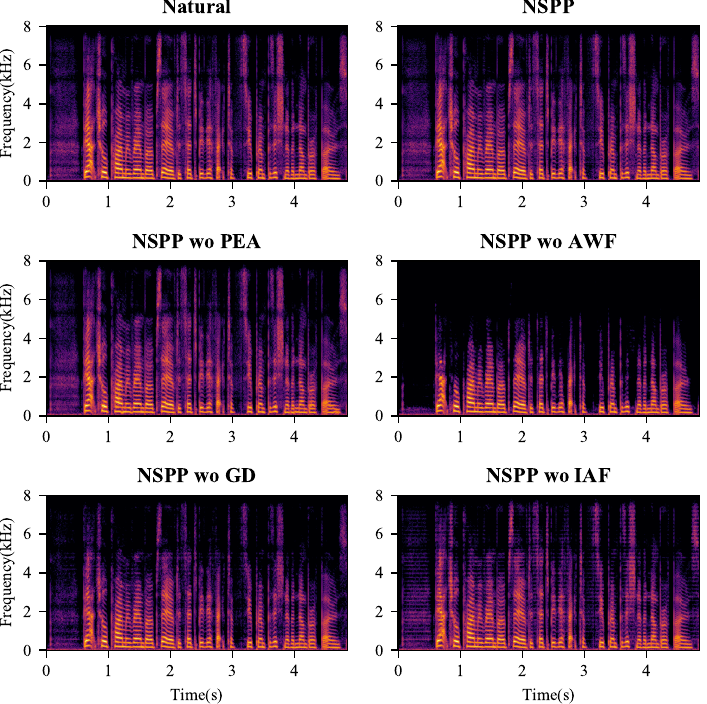}
    \caption{A comparison among the spectrograms of the natural speech and speeches generated by \textbf{NSPP}, \textbf{NSPP wo PEA}, \textbf{NSPP wo AWF}, \textbf{NSPP wo GD} and \textbf{NSPP wo IAF} for the analysis-synthesis task.}
    \label{fig: Spectrogram}
\end{figure}

\begin{figure}
    \centering
    \includegraphics[height=\linewidth]{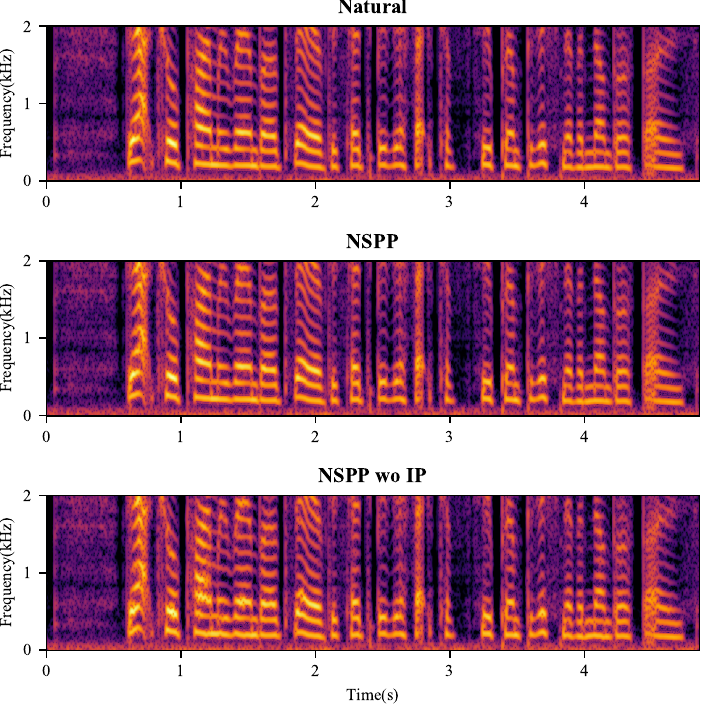}
    \caption{A comparison among the low-frequency (0$\sim$2000Hz) spectrograms of the natural speech and speeches generated by \textbf{NSPP} and \textbf{NSPP wo IP} for analysis-synthesis task.}
    \label{fig: IP_destory}
\end{figure}

We also utilized SNR and F0-RMSE as objective metrics here for evaluating the reconstructed speech.
Regarding the subjective evaluations, we conducted ABX preference tests on the Amazon Mechanical Turk platform to compare the differences between the \textbf{NSPP} and its ablated variants.
The objective and subjective results are listed in Table \ref{tab_results_ablation} and Figure \ref{fig: ABX}, respectively.
Additionally, we also provided the spectrograms of the speeches generated by the \textbf{NSPP} and its ablated variants in Figures \ref{fig: Spectrogram} and \ref{fig: IP_destory} for visual analysis.

\begin{table}
\centering
    \caption{Objective evaluation results among \textbf{NSPP} and its ablated variants for the analysis-synthesis task.}
    %\resizebox{8.8cm}{1.05cm}{
    \begin{tabular}{c | c c}
        \hline
        \hline
        & SNR(dB)$\uparrow$ & F0-RMSE(cent)$\downarrow$\\
        \hline
        \textbf{NSPP} & 8.26 & \textbf{10.0}\\
        \hline
        \textbf{NSPP wo PEA} & 4.65 & 36.9\\
        \textbf{NSPP wo AWF} & 8.51 & 12.0\\
        \textbf{NSPP wo IP} & 4.95 & 21.2\\
        \textbf{NSPP wo GD} & \textbf{8.95} & 10.1\\
        \textbf{NSPP wo IAF} & 8.69 & 12.1\\
        \hline
        \hline
    \end{tabular}%}
\label{tab_results_ablation}
\end{table}

As expected, we can see that the \textbf{NSPP} outperformed the \textbf{NSPP wo PEA} significantly ($p<0.01$) by analyzing both objective and subjective results listed in Table \ref{tab_results_ablation} and Figure \ref{fig: ABX}, respectively.
Specifically, the speech reconstructed by the \textbf{NSPP wo PEA} exhibited annoying loud noise similar to electric current, which significantly affected the sense of hearing due to the imprecise phase prediction.
By comparing the spectrograms of the speeches generated by the \textbf{NSPP} and \textbf{NSPP wo PEA} in Figure \ref{fig: Spectrogram}, we can find that the spectrogram of the \textbf{NSPP wo PEA} was a little blurred.
One possible reason is that it was difficult for neural networks without the parallel estimation architecture to restrict the range of predicted phases, leading to a failure of anti-wrapping losses.
These results indicated that the parallel estimation architecture was essential to wrapped phase prediction.

By comparing \textbf{NSPP} and \textbf{NSPP wo AWF} in Table \ref{tab_results_ablation}, the SNR of the \textbf{NSPP wo AWF} was even higher and the F0-RMSE of the \textbf{NSPP wo AWF} was comparable to that of the \textbf{NSPP}.
However, the subjective results in Figure \ref{fig: ABX} indicated that the \textbf{NSPP} outperformed the \textbf{NSPP wo AWF} significantly ($p<0.01$) in terms of speech quality, which proved that the anti-wrapping function was helpful for avoiding the error expansion issue.
As shown in Figure \ref{fig: Spectrogram}, the high-frequency energy of the speech reconstructed by the \textbf{NSPP wo AWF} was completely suppressed, resulting in an extremely dull listening experience.
This may be the reason for the poor ABX scores.
Interestingly, there was no obvious mispronunciation or F0 distortion in the speech reconstructed by the \textbf{NSPP wo AWF} (F0-RMSE=12.0 cent, comparable to that of \textbf{NSPP}).
The above experimental results also confirmed that the absence of high-frequency components had little effect on the SNR metric.

For the three losses, removing any loss led to poor performance of the model.
However, each loss played a very different role.
Removing $\mathcal L_{IP}$ (i.e., \textbf{NSPP wo IP}) led to a sharp drop in all objective metrics in Table \ref{tab_results_ablation}.
Regarding the ABX test results in Figure \ref{fig: ABX}, the subjective difference between the \textbf{NSPP} and \textbf{NSPP wo IP} was slightly insignificant ($p$ was slightly larger than 0.01).
However, we found that the reconstructed speech quality of the \textbf{NSPP wo IP} indeed degraded.
Figure \ref{fig: IP_destory} shows the low-frequency F0 and harmonic details of the spectrograms of the \textbf{NSPP} and \textbf{NSPP wo IP}.
Obviously, the speech reconstructed by the \textbf{NSPP wo IP} exhibited few low-frequency spectrum corruption issues (see the range of 0.5$\sim$2 seconds in Figure \ref{fig: IP_destory}), resulting in F0 and harmonic structure distortion and blurry pronunciation.
This is also the reason why the F0-RMSE of the \textbf{NSPP wo IP} was relatively poor.
However, removing $\mathcal L_{GD}$ (i.e., \textbf{NSPP wo GD}) and $\mathcal L_{IAF}$ (i.e., \textbf{NSPP wo IAF}) did not cause significant deterioration on all objective metrics in Table \ref{tab_results_ablation}.
Although the subjective difference between the \textbf{NSPP} and \textbf{NSPP wo GD} was slightly insignificant ($p$ was slightly larger than 0.01) in Figure \ref{fig: ABX}, we can see from Figure \ref{fig: Spectrogram} that the \textbf{NSPP wo GD} attenuated the overall spectral energy of the reconstructed speech, resulting in a mild dull listening experience.
As shown in Figure \ref{fig: ABX}, removing $\mathcal L_{IAF}$ (i.e., \textbf{NSPP wo IAF}) led to a significant subjective performance degradation ($p<0.01$), manifested in the presence of obvious spectral horizontal stripes in the reconstructed speech (see Figure \ref{fig: Spectrogram}), causing annoying loud noise.
Interestingly, the removal of $\mathcal L_{GD}$ and $\mathcal L_{IAF}$ did not destroy the F0 and harmonic structure, nor did it lead to mispronunciation (their F0-RMSEs were comparable to that of \textbf{NSPP}).

In conclusion, all ablated elements were indispensable for our proposed neural speech phase prediction model.
The parallel estimation architecture and IP loss prevented the destruction of the F0 and spectral structure of speech.
The anti-wrapping function and GD loss avoided the attenuation of high-frequency energy and dull hearing of speech.
The IAF loss suppressed the appearance of loud noise caused by spectral horizontal lines.

\section{Conclusion}
\label{sec: Conclusion}

In this paper, we have proposed a novel neural speech phase prediction model, which utilizes a residual convolutional network along with a parallel estimation architecture to directly predict the wrapped phase spectra from input amplitude spectra.
The parallel estimation architecture is a key module which consists of two parallel linear convolutional layers and a phase calculation formula, strictly restricting the output phase values to the principal value interval.
The training criteria of the proposed model are to minimize a combination of the instantaneous phase loss, group delay loss and instantaneous angular frequency loss, which are all activated by an anti-wrapping function to avoid the error expansion issue caused by phase wrapping.
The anti-wrapping function should possess three properties, i.e., parity, periodicity and monotonicity.
Low-latency streamable phase prediction is also achieved with the help of causal convolutions and knowledge distillation training strategies.
Experimental results show that the proposed model outperforms the GLA, RAAR and von Mises distribution DNN-based phase prediction methods for both analysis-synthesis and specific speech generation tasks (i.e., the BWE and SS) in terms of phase prediction precision, efficiency and robustness.
The proposed model is significantly faster in generation speed than HiFi-GAN-based waveform reconstruction method, while also having the same synthesized speech quality.
%and HiFi-GAN-based waveform reconstruction method for both analysis-synthesis and specific speech generation tasks (i.e., the BWE and SS) in terms of phase prediction precision, efficiency, latency and robustness.
%When initially applied to specific speech generation tasks, the proposed model shows better robustness compared with iterative algorithms.
Besides, the proposed model is easy to implement and also exhibits a fast training speed.
%Ablation studies demonstrate that the parallel estimation architecture, anti-wrapping function and three losses are all useful.
Integrating the neural speech phase prediction model to more end-to-end speech generation tasks will be the focus of our future work.

\bibliographystyle{IEEEtran}
\bibliography{mybib}
\end{document}